\documentclass[acmsmall,screen]{acmart}
\AtBeginDocument{%
  }

\setcopyright{cc}
\setcctype{by}
\acmDOI{10.1145/3808302}
\acmYear{2026}
\acmJournal{PACMPL}
\acmVolume{10}
\acmNumber{PLDI}
\acmArticle{224}
\acmMonth{6}
\acmSubmissionID{pldi26main-p376-p}
\received{2025-11-13}
\received[accepted]{2026-04-03}



\usepackage{algorithm}
\usepackage{algpseudocode}
\usepackage{amsmath}
\usepackage{amssymb}
\usepackage{amsthm}
\usepackage{array}
\usepackage{caption}
\usepackage{float}
\usepackage{graphicx}
\usepackage{listings}
\usepackage{listings-rust}
\usepackage{makecell}
\usepackage{mathrsfs}
\usepackage{semantic}
\usepackage{stmaryrd}
\usepackage{subcaption}
\usepackage{tikz}
\usepackage{wrapfig}
\usepackage{xcolor}
\usepackage{zi4}

\usetikzlibrary{quotes, graphs}

\usepackage{xspace}
\newcommand{\egg}{\texttt{egg}\xspace}

\newcommand{\egraphs}{\mbox{e-graphs}\xspace}
\newcommand{\egraph}{\mbox{e-graph}\xspace}

\newcommand{\Egraphs}{\mbox{E-graphs}\xspace}
\newcommand{\eclass}{\mbox{e-class}\xspace}

\newcommand{\denote}[1]{\ensuremath{\llbracket #1 \rrbracket}}

\theoremstyle{remark}

\definecolor{codegreen}{rgb}{0,0.6,0}
\definecolor{codegray}{rgb}{0.5,0.5,0.5}
\definecolor{codepurple}{rgb}{0.58,0,0.82}
\definecolor{backcolour}{rgb}{0.95,0.95,0.92}

\lstdefinestyle{mystyle}{
    backgroundcolor=\color{backcolour},   
    commentstyle=\color{codegreen},
    keywordstyle=\color{magenta},
    numberstyle=\tiny\color{codegray},
    stringstyle=\color{codepurple},
    basicstyle=\ttfamily\footnotesize,
    breakatwhitespace=false,         
    breaklines=true,                 
    captionpos=b,                    
    keepspaces=true,                 
    numbers=left,                    
    numbersep=5pt,                  
    showspaces=false,                
    showstringspaces=false,
    showtabs=false,                  
    tabsize=2
}

\newcolumntype{x}[1]{>{\centering\arraybackslash\hspace{0pt}}p{#1}}

\newif\ifcomments
\ifcomments
    \providecommand{\alvin}[1]{{\color{brown}{alvin: #1 }}}
    \newcommand{\mw}[1]{{\color{red}{MW: #1}}}
    \newcommand{\russel}[1]{{\color{blue}{russel: #1}}}
\else
    \providecommand{\alvin}[1]{}
    \newcommand{\mw}[1]{}
    \newcommand{\russel}[1]{}
\fi

\begin{document}
\title{Optimism in Equality Saturation}
\author{Russel Arbore}
\email{russel@berkeley.edu}
\orcid{0009-0003-8700-0846}
\affiliation{%
  \institution{University of California, Berkeley}
  \city{Berkeley}
  \state{California}
  \country{USA}
}
\author{Alvin Cheung}
\email{akcheung@cs.berkeley.edu}
\orcid{0000-0001-6261-6263}
\affiliation{%
  \institution{University of California, Berkeley}
  \city{Berkeley}
  \state{California}
  \country{USA}
}
\author{Max Willsey}
\email{mwillsey@berkeley.edu}
\orcid{0000-0001-8066-4218}
\affiliation{%
  \institution{University of California, Berkeley}
  \city{Berkeley}
  \state{California}
  \country{USA}
}

\begin{abstract}

Equality saturation
 is a 
 program optimization technique based 
 on non-destructive rewriting
 and 
 a form of abstract interpretation called e-class analysis.
Existing e-class analyses 
 are pessimistic
 and therefore typically imprecise when analyzing cyclic programs,
 such as those in SSA form.
We show that a straightforward optimistic variant of e-class analysis can result in unsoundness, due to a subtlety in how e-graphs represent programs.
We propose an abstract interpretation algorithm that circumvents this issue and can optimistically analyze e-graphs during equality saturation.
This results in a unified algorithm for optimistic analysis and non-destructive rewriting.
We implement a prototype abstract interpreter
 and equality saturation tool for SSA programs.
Our tool exhibits precision improvements over pure abstract interpretation (without rewriting) and pessimistic e-class analysis on example programs. Additionally, its performance is comparable to existing abstract interpretation and e-class analysis techniques.
\end{abstract}

\maketitle

\section{Introduction}

Optimizing compilers transform an input program into a ``better'' program. Most compilers implement a transformation-based approach---an input program is modified by a sequence of separate passes. In this regime, the order of the individual passes matters, because each pass destroys the previously known program (this is often referred to as the phase ordering problem).

\paragraph{Equality Saturation}
Equality saturation is an alternate approach to performing compilation, where transformations are implemented as term rewrites and a data structure called an e-graph keeps track of all intermediate programs and known equivalences between them \cite{e_peg}. This approach enables \emph{non-destructive} rewriting, which bypasses the phase ordering problem. 
Recent implementations of equality saturation also include \emph{e-class analysis}, which is a form of abstract interpretation~\cite{egg, egraphs_plus_ai}.
Rewrites and analyses cooperate:
 rewrites can be conditioned on analysis facts,
 and equalities from rewriting combine analysis facts
 into more precise ones~\cite{egraphs_plus_ai}.
These systems have been applied in many areas, including
 floating point accuracy \cite{herbie}, 
 circuit synthesis~\cite{rover_hw},
 tensor and linear algebra~\cite{tensat, spores}, 3D CAD \cite{cad}, 
 and imperative program compilation \cite{e_peg, aegraphs}.

\paragraph{Optimism}
Many interesting programs contain loops, which often correspond to cycles in data flow. A strict subset of program analyses, called ``optimistic'' analyses in the compilers literature, are capable of precisely analyzing cyclic program representations, as opposed to ``pessimistic'' analyses, which use bottom-up reasoning and therefore struggle to reason about loops.
Optimistic analyses, as their name suggests, optimistically assume a potentially unsound analysis fact about program fragments in a cycle and then later refine the analysis \cite{combining_analyses, combining_program_improvers}. 

\paragraph{Optimism in Equality Saturation}
Incorporating optimistic analyses into equality saturation has not been achieved previously. Existing e-class analyses are pessimistic, meaning cyclic programs are not precisely analyzed. 
Additionally, performing optimistic analyses on e-graphs after rewriting is fraught; \emph{we show that the straightforward approach leads to unsoundness}. 
Additionally, as optimistic analyses are not incrementally sound, optimistic analyses cannot be interleaved with rewriting in equality saturation, as rewrites are not necessarily revocable in e-graphs. 
Prior work in equality saturation has identified a specific need for optimism: de-duplicating isomorphic cycles in an e-graph \cite{e_peg, co_egraphs, omelets_need_onions}.
A technique from traditional compilers,
 optimistic global value numbering, would solve this problem \cite{gvn, scc_gvn, combining_analyses},
 but is not applicable due to the aforementioned difficulties.

We incorporate optimistic analyses into an equality saturation system for the first time.
We describe a prototype combined equality saturation engine and
 abstract interpreter for imperative programs,
 based on a Static Single Assignment (SSA) program representation with novel semantics.
Our prototype analyzes example programs more precisely 
 than standard abstract interpretation, pessimistic e-class analysis, or compilation with \texttt{gcc} or \texttt{clang}.
It also presents a solution to the cycle de-duplication
 problem identified in prior work.
In summary, our contributions are as follows:
\begin{itemize}
    \item We identify the core issue preventing optimism in equality saturation---equality saturation creates ill-formed represented graphs which poison optimistic analyses (Section~\ref{sec:challenges}).
    \item We propose a SSA form program representation that 1) can be easily embedded into an e-graph and 2) has a simple semantics amenable to abstract interpretation (Section~\ref{sec:semantics_and_ai_ssa}).
    \item We describe optimistic analyses over SSA programs embedded in e-graphs---discovered equalities make analyses more precise, but also create ill-formed represented graphs. We propose an abstract interpretation algorithm that computes an abstraction that is both sound and optimistic. We prove that this algorithm always computes a sound analysis of the well-formed represented graphs in an e-graph and study its complexity (Section~\ref{sec:ai_over_ssa_egraphs}).
    \item We build a prototype program optimizer in Rust that combines equality saturation and optimistic analyses and evaluate it on example programs requiring both rewriting and optimistic analysis to fully optimize; existing techniques cannot fully optimize these examples, while our tool can. We also study the cost of our algorithm on randomly generated programs, compared against standard abstract interpretation and e-class analysis. (Sections~\ref{sec:ai_plus_eqsat}~and~\ref{sec:implementation_and_examples}).
\end{itemize}



\section{Background}


\subsection{Equality Saturation}
\label{sec:background_eqsat}

Equality saturation \cite{e_peg} 
performs non-destructive rewriting
over the \egraph data structure \cite{nelson1980}. Several implementations exist, the two most popular being \texttt{egg} \cite{egg} and \texttt{egglog} \cite{egglog}.

\subsubsection{E-Graphs}

An \emph{e-graph} stores an equivalence (sometimes congruence) relation of terms.

\begin{definition}[E-Graphs]
\label{def:background_e_graphs}
An e-graph is pair of finite sets of \emph{e-nodes} $\mathcal{N}$ and \emph{e-classes} $\mathcal{C}$ where:
\begin{itemize}
    \item An e-node $n \in \mathcal{N}$ is a \emph{function symbol} $f$ of arity $k$ and a tuple of e-classes $(i_1,\ldots,i_k) \in C^k$. 
    An e-node ``is'' a function $f$ when $f$ is its symbol.
    The $j$th input class is written as $n_j$.
    \item There is a surjection $[\cdot] \in \mathcal{N} \rightarrow \mathcal{C}$ that defines e-class membership. $[\cdot]^{-1} \in \mathcal{C} \rightarrow \mathcal{P}(\mathcal{N})$ 
    gives the pre-image of an e-class under $[\cdot]$.
\end{itemize}
\end{definition}

Prior work defines "represented terms" to describe programs existing in an e-graph \cite{egg}. We instead define \emph{represented graphs}, since we will use e-graphs to represent cyclic terms in this paper.

\begin{definition}[Cyclic Terms]
\label{def:background_cyclic_terms}
A cyclic term is a pair of finite sets of nodes $\mathcal{V}$ and ids $\mathcal{I}$ where:
\begin{itemize}
    \item A node $v \in \mathcal{V}$ is a \emph{function symbol} $f$ of arity $k$ and a tuple of ids $(i_1,\ldots,i_k) \in \mathcal{I}^k$. A node ``is'' a function $f$ when $f$ is its symbol.
    \item There is a bijection $v \in \mathcal{I} \rightarrow \mathcal{V}$.
\end{itemize}
For brevity, we conflate $\mathcal{V}$ and $\mathcal{I}$ (and simply refer to $\mathcal{V}$)---for example, we write the $j$th input to a node $n$ as $n_j$ and say $n_j \in \mathcal{V}$ (while technically we must apply $v$ to map from the input id to a node). The graph may contain cycles. We often call cyclic terms just ``graphs''.
\end{definition}

\begin{definition}[Represented Graphs]
\label{def:background_represented_graphs}
A cyclic term $\mathcal{V}$ is a represented graph of an e-graph $(\mathcal{N}, \mathcal{C})$ when there is a map $m \in \mathcal{V} \rightarrow \mathcal{N}$ such that for all nodes $v \in \mathcal{V}$, $v$ and $m(v)$ have the same function symbol, $\forall i \in \mathbb{Z}, 1 \le i \le k \implies m(v)_i = [m(v_i)]$, where $k$ is the arity of $v$'s function symbol, and $[\cdot] \circ m$ is a surjection. In other words, $m$ is a homomorphism from the cyclic term into the e-graph preserving dependency structure up to equivalence and all e-classes are mapped into\footnote{Technically, prior definitions of represented terms do not require that all e-classes are represented. We use this requirement to simplify some definitions in Section~\ref{sec:ai_over_ssa_egraphs}, but represented graphs can be understood without this requirement.}. 
\end{definition}

Figure~\ref{fig:egraph-example} shows three example e-graphs. The e-nodes and e-classes of an e-graph are mutually recursive---e-nodes are members of e-classes and e-nodes have e-classes as inputs rather than e-nodes, since it doesn't matter which e-node in an e-class is being referred to (they're all equivalent).

\begin{figure}
    \centering
    \begin{subfigure}[t]{0.32\textwidth}
        \centering
        \begin{tikzpicture}
            \graph[grow left, branch down] { 
                {nab[as=$-^{\textcolor{red}{e}}$], nba[as=$-^{\textcolor{red}{f}}$]} <- {ab[as=$+^{\textcolor{red}{c}}$], ba[as=$+^{\textcolor{red}{d}}$]} <- {a[as=$x^{\textcolor{red}{a}}$], b[as=$y^{\textcolor{red}{b}}$]};
                {b, a} -> {ab, ba};
            };
        \end{tikzpicture}
        \caption{Initial e-graph representing $-(x + y)$ and $-(y + x)$.}
        \label{fig:egraph-example-first}
    \end{subfigure}
    \hfill
    \begin{subfigure}[t]{0.32\textwidth}
        \centering
        \begin{tikzpicture}
            \graph[grow left, branch down] { 
                {nab[as=$-^{\textcolor{red}{e}}$], nba[as=$-^{\textcolor{red}{f}}$]} <- {ab[as=$+^{\textcolor{red}{c}}$], ba[as=$+^{\textcolor{red}{c}}$]} <- {a[as=$x^{\textcolor{red}{a}}$], b[as=$y^{\textcolor{red}{b}}$]};
                {b, a} -> {ab, ba};
                ab --[red,dashed,thick] ba;
            };
        \end{tikzpicture}
        \caption{E-Graph after applying the rewrite $x + y \Rightarrow y + x$.}
        \label{fig:egraph-example-second}
    \end{subfigure}
    \hfill
    \begin{subfigure}[t]{0.32\textwidth}
        \centering
        \begin{tikzpicture}
            \graph[grow left, branch down] { 
                {nab[as=$-^{\textcolor{red}{e}}$], nba[as=$-^{\textcolor{red}{e}}$]} <- {ab[as=$+^{\textcolor{red}{c}}$], ba[as=$+^{\textcolor{red}{c}}$]} <- {a[as=$x^{\textcolor{red}{a}}$], b[as=$y^{\textcolor{red}{b}}$]};
                {b, a} -> {ab, ba};
                ab --[red,dashed,thick] ba;
                nab --[red,dashed,thick] nba;
            };
        \end{tikzpicture}
        \caption{E-Graph after rebuilding.}
        \label{fig:egraph-example-third}
    \end{subfigure}
    \caption{
    Example e-graphs during equality saturation. 
    Symbols are e-nodes, 
     solid edges connect e-nodes to arbitrary e-nodes in their input e-classes,
     and 
     dashed \textcolor{red}{red} edges connect e-nodes in the same e-class, following the convention from \cite{e_peg}. We super-script e-nodes with an identifier (also in \textcolor{red}{red}) for their e-class.
     Note that we connect 
     e-nodes \textleftarrow{} input e-classes, 
     indicating the flow of data, as in \cite{sea_of_nodes}.
    }
    \label{fig:egraph-example}
\end{figure}

\subsubsection{Batched Rewriting and Rebuilding}

A typical equality saturation workflow has three steps:

\begin{enumerate}
    \item 
    The e-graph is seeded with an initial graph (the graph we are interested in optimizing). 
    \item
    Rewrites rules are repeatedly applied to the e-graph.
    This may reach a fixpoint (called saturation), but is not guaranteed to---applications often stop rewriting after a timeout \cite{semantic_eqsat}.
    \begin{enumerate}
        \item 
        Rewrites query the e-graph using the \emph{e-matching}
         procedure to find terms represented by the e-graph 
         matching the left hand side of the rewrite, like the $x + 0$ in $x + 0 \Rightarrow x$.
        \item
        Matches yield substitutions, which are used to instantiate the right hand side pattern---these new terms are inserted and asserted equal with matched terms.
        \item
         Discovered equalities may imply further equalities by congruence, so \emph{rebuilding} is run to discover these new equalities explicitly.
    \end{enumerate}
    \item
    \emph{Extraction} picks a graph (usually a tree or DAG) represented by the e-graph that is optimized with respect to some cost model \cite{e_peg, egg, smoothe, eboost, treewidth-extract}. Extraction is an e-class analysis \cite{egg} and prior work has encountered a similar problem to the one we describe in Section~\ref{sec:challenges} \cite{tensat}.
\end{enumerate}


\subsubsection{E-Class Analysis}
\label{sec:background_eclass_analysis}

The \emph{e-class analysis} framework adds abstract interpretation capabilities into equality saturation, enabling the safe application of rewrite rules that are conditioned on some analysis result \cite{egg, egraphs_plus_ai}.
E-class analysis associates a semi-lattice element with each e-class and propagates facts in a bottom-up fashion. Facts are associated with e-classes, rather than e-nodes, because all e-nodes in an e-class are known to be equal---a sound analysis for any e-node in the e-class must be sound for all of the e-nodes. \texttt{egglog} uses a Datalog-like architecture to associate multiple semi-lattice facts with each e-class, but the fundamental mechanism is the same \cite{egglog}. Facts derived for e-nodes inside the same e-class can be combined to increase precision---since rewrites create larger e-classes, rewriting and e-class analysis are mutually beneficial \cite{egg, egglog, egraphs_plus_ai}.

Consider the following example of an interval e-class analysis
 from Coward et.\ al.~\cite{egraphs_plus_ai}.
Let our language be arithmetic expressions,
 and the analysis domain be the set of intervals over the reals.
Consider the following expressions over
  variables $x$ and $y$, where $x \in [0,1]$ and $y \in [1, 2]$.
\[
\frac{x-y}{x+y} \in [-2, 0] \hspace{3em}
\frac{2x}{x+y} - 1 \in [-1, 1]
\]

Since the variables $x$ and $y$ have known intervals,
 the intervals for the terms can be derived as $[-2, 0]$ and $[-1, 1]$.
If rewriting discovers that the two 
 above expressions are equivalent (they are),
 the e-classes will be merged,
 and the fact for the new e-class
 will be their \emph{intersection}
 (if two equivalent terms have intervals,
 they both must lie in both intervals).
Thus, the new fact for the merged \eclass
 will be $[-1, 0]$,
 which is more precise than either of the original intervals.

\subsubsection{Cycles in E-Graphs}


Rewrites can create cycles when a term is discovered equivalent to one of its sub-terms. In prior work, these cycles correspond to infinite families of acyclic terms. 
In this work, we embed cyclic terms into the e-graph, which requires extending the notion of representation to graphs (Definition~\ref{def:background_represented_graphs}). However, \emph{this does not mechanically change the e-graph}.

\subsection{Abstract Interpretation}
\label{sec:background_ai}


Abstract interpretation is a framework for approximating the behavior of programs, usually using lattice structures \cite{abstract_interpretation, systematic_program_analysis}. ``Abstractions'' characterize what sets of concrete executions are possible in a particular program \cite{mine_dbms, mine_graph_based, mine_octagon, abstract_interpretation, pentagons, singh_octagon, polyhedra, fast_polyhedra, karr, tvpi, labeled_union_find, tristate}. Often, abstractions characterize what concrete values a program variable may store at some point during program execution---these are called non-relational abstractions, and include intervals \cite{abstract_interpretation} and known-bits \cite{tristate}. Abstractions can also characterize what concrete values a tuple of program variables may store simultaneously---these are called relational abstractions, and include difference bounds \cite{mine_dbms}, octagons \cite{mine_octagon, singh_octagon}, pentagons \cite{pentagons}, polyhedra \cite{polyhedra, fast_polyhedra}, Karr's domain \cite{karr}, two variables per inequality \cite{tvpi}, and equality \cite{alien_expressions}.

Typically, abstract interpretation is described in terms of a \emph{Galois connection} between a \emph{concrete lattice} $\Sigma$ and an \emph{abstract lattice} $\Sigma^\#$. Concretization and abstraction functions $\gamma \in \Sigma^\# \rightarrow \Sigma$ and $\alpha \in \Sigma\rightarrow\Sigma^\#$ form a Galois connection when for all $s \in \Sigma$ and $s^\# \in \Sigma^\#$, $s \sqsubseteq \gamma(s^\#) \iff \alpha(s) \sqsubseteq^\# s^\#$. An abstraction $s^\# \in \Sigma^\#$ is a \emph{sound} over-approximation of $s \in \Sigma$ when $s \sqsubseteq \gamma(s^\#)$. We can lift a function $f \in \Sigma \rightarrow \Sigma$ to operate on $\Sigma^\#$: $\alpha \circ f \circ \gamma \in \Sigma^\# \rightarrow \Sigma^\#$. $\gamma$ and $\alpha$ may be incomputable---an \emph{abstract transformer} $f^\# \in \Sigma^\# \rightarrow \Sigma^\#$ soundly over-approximates $f$ when for all $s^\# \in \Sigma^\#$, $f (\gamma(s^\#)) \sqsubseteq \gamma(f^\#(s^\#))$. We use the $\#$ superscript notation to refer to the abstract counterpart of a concrete object.

An abstract interpretation can be expressed as a set of equations describing states at program locations as abstract transformers applied to states at predecessor locations. These equations may be mutually dependent. If the abstract transformers are sound, any fixpoint solution to the equations is sound with respect to the program semantics \cite{abstract_interpretation}. For some domains, infinite ascending or descending chains may prevent the computation of least or greatest fixpoints, respectively. A sound analysis can be computed by using widening or narrowing, respectively, to compute a \emph{post}-fixpoint instead (abstraction where further iteration of the equations yields a strictly decreased abstraction).

\begin{figure}
    \centering
    \hfill
    \begin{subfigure}[c]{0.3\textwidth}
        \centering
        \begin{lstlisting}[language=Rust]
let x = 1;
while 1 {
    x = x + (1 * 5);
}\end{lstlisting}
        \caption{A simple program with a loop.}
        \label{fig:ssa-example-program}
    \end{subfigure}
    \hfill
    \begin{subfigure}[c]{0.3\textwidth}
        \centering
    \begin{tikzpicture}
        \graph[no placement] { 
            phi1[as=$\phi_v^{\textcolor{red}{c}}$,at={(4,0)}];
            pp[as=$+^{\textcolor{red}{d}}$,at={(5, 0)}];
            1[as=$1^{\textcolor{red}{a}}$,at={(5, 1)}];
            t[as=$*^{\textcolor{red}{e}}$,at={(6,0)}];
            5[as=$5^{\textcolor{red}{b}}$,at={(6,1)}];
            1 -> phi1;
            t -> pp;
            pp ->[bend right = 30] phi1;
            phi1 ->[bend right = 30] pp;
            1 -> t;
            5 -> t;
        };
    \end{tikzpicture}
        \caption{DFG of the program.}
        \label{fig:ssa-example-graph}
    \end{subfigure}
    \hfill
    \begin{subfigure}[c]{0.3\textwidth}
        \centering
    \begin{tikzpicture}
        \graph[no placement] { 
            s[as=$s$,at={(0,0)}];
            v[as=$v$,at={(1, 0)}];
            s ->["\textcolor{red}{\scriptsize $a$}"] v;
            v ->["\textcolor{red}{\scriptsize $a$}" above, inner sep=8pt, looseness=8,min distance=5mm,in=30,out=-30] v;
        };
    \end{tikzpicture}
        \caption{CFG of the program.}
        \label{fig:ssa-example-cfg}
    \end{subfigure}
    \hfill
    \caption{A program in pseudo-code and as a SSA program. The nodes in the DFG are super-scripted by identifiers in \textcolor{red}{red}. The CFG contains two nodes: $s$, which is the entry point, and $v$, representing the loop body. $\phi$ nodes are annotated with CFG vertices. CFG edges are annotated by predicate values that guard control flow.}
    \label{fig:ssa-graph-example}
\end{figure}

\subsection{Static Single Assignment Form}
\label{sec:background_ssa}

\emph{Static Single Assignment} (SSA) form is a category of program representation where 1) every variable has a single definition and 2) every definition is executed before its uses (which is called \emph{dominance}). Variables in SSA form are often instead called \emph{values} to emphasize that they do not change and are unambiguously defined. A special $\phi$ instruction is used to join data flow at control flow joins \cite{ssa_book}. 

Some SSA form representations use graphs to represent data flow, where nodes represent operators and edges represent dependencies---SSA values are identified by nodes, and $\phi$ nodes are used to join results at control flow joins. Prior works have multiple names and exact formulations for these graphs. One form is the ``sea-of-nodes'' graph representing data and control flow \cite{sea_of_nodes, semantic_sea_of_nodes}. The data flow graph has also been called the ``global value graph'' (``SSA graphs'' are described as combining global value graphs and control flow graphs) \cite{ssa_translation_ai}. We separate the data flow and control flow portions of a program into a data flow graph (DFG) and a control flow graph (CFG). Figure~\ref{fig:ssa-graph-example} shows an example program, a corresponding DFG, and a corresponding CFG. 

Not all DFGs can be in SSA form; if there is a dependency cycle consisting of non-$\phi$ operations, there is no order in which the operations can be evaluated (in every serialization of the cycle, at least one definition will not dominate all of its uses). 
Additionally, it only makes sense to talk about a DFG being in SSA form with respect to some CFG, as the CFG defines dominance among $\phi$ nodes.
We say that a DFG that is in SSA form (with respect to some CFG) is ``well-formed''. 

\section{Challenges in Abstract Interpretation over E-Graphs}
\label{sec:challenges}

We describe in detail why performing abstract interpretation over e-graphs, specifically in the presence of cycles, is challenging. We identify how existing e-class analysis implementations circumvent this issue and argue that this circumvention is insufficient for future use cases.

Recall from Section~\ref{sec:background_ai} that we can express an analysis of a program as a set of equations over an abstract lattice. We can express an e-class analysis over an e-graph in a similar fashion. Additionally, recall from Section~\ref{sec:background_eqsat} that e-class analysis associates an analysis fact with each e-class in an e-graph. Thus, these equations will describe the known analysis fact for each e-class. Each e-node can be abstractly interpreted according to 1) its function symbol and 2) its input e-classes. Additionally, the abstractions of multiple e-nodes in an e-class can be combined via the meet operator to obtain an abstraction of the e-class. If $f$ is the function symbol of an e-node, call $f^\#$ a corresponding abstract transformer. Then, the equation for the abstraction $c^\#$ describing an e-class $c$ is: $$c^\#=\bigsqcap_{n \in [c]^{-1}} f^\#(n_1^\#,\ldots,n_k^\#)$$ 

\begin{figure}
    \centering
    \hfill
    \begin{subfigure}[c]{0.21\textwidth}
        \centering
    \begin{tikzpicture}
        \graph[no placement] { 
            1[as=$1^{\textcolor{red}{a}}$,at={(0, 1)}];
            t[as=$*^{\textcolor{red}{b}}$,at={(0.5,0)}];
            5[as=$5^{\textcolor{red}{b}}$,at={(1,1)}];
            1 -> t;
            5 -> t;
            5 --[red,bend left=50,dashed,thick] t;
        };
    \end{tikzpicture}
        \caption{E-graph representing $1 * 5$, with the rewrite $1 * x \Rightarrow x$.}
        \label{fig:conventional_egraphs_and_eqs_egraph}
    \end{subfigure}
    \hfill
    \begin{subfigure}[c]{0.25\textwidth}
    \centering
    \begin{align*}
        \textcolor{red}{a}^\# &= [1,1] \\
        \textcolor{red}{b}^\# &= [5,5] \sqcap (\textcolor{red}{a}^\# * \textcolor{red}{b}^\#)
    \end{align*}
        \caption{Analysis equations for the e-graph in Figure~\ref{fig:conventional_egraphs_and_eqs_egraph}.}
        \label{fig:conventional_egraphs_and_eqs_eqs}
    \end{subfigure}
    \hfill
    \begin{subfigure}[c]{0.19\textwidth}
        \centering
    \begin{tikzpicture}
        \graph[no placement] { 
            1[as=$1^{\textcolor{red}{a}}$,at={(0, 1)}];
            t[as=$*^{\textcolor{red}{b}}$,at={(0.5,0)}];
            1 -> t;
        };
        \draw[->] (t) to [out=330,in=30,looseness=3] (t);
    \end{tikzpicture}
        \caption{Ill-formed graph that is represented in e-graph of Figure~\ref{fig:conventional_egraphs_and_eqs_egraph}.}
        \label{fig:conventional_egraphs_and_eqs_ill_formed_graph}
    \end{subfigure}
    \hfill
    \begin{subfigure}[c]{0.25\textwidth}
    \centering
    \begin{align*}
        \textcolor{red}{a}^\# &= [1,1] \\
        \textcolor{red}{b}^\# &= \textcolor{red}{a}^\# * \textcolor{red}{b}^\# 
    \end{align*}
        \caption{Analysis equations for the graph in Figure~\ref{fig:conventional_egraphs_and_eqs_ill_formed_graph}.}
        \label{fig:conventional_egraphs_and_eqs_ill_formed_eqs}
    \end{subfigure}
    \hfill
    \caption{E-class analysis (with the interval lattice) of a simple e-graph.}
    \label{fig:conventional_egraphs_and_eqs}
\end{figure}

This formulation is standard from prior e-class analysis literature \cite{egg, egraphs_plus_ai}. Let us now consider applying e-class analysis to a specific example. Figures~\ref{fig:conventional_egraphs_and_eqs_egraph}~and~\ref{fig:conventional_egraphs_and_eqs_eqs} show an e-graph and its e-class analysis equations. We use the interval lattice to analyze a very simple e-graph that is representative of structures seen in existing applications of e-graphs. There are two fixpoints to this set of equations: $\textcolor{red}{a}^\# = [1,1], \textcolor{red}{b}^\# = [5,5]$ (the greatest fixpoint), and $\textcolor{red}{a}^\# = [1,1], \textcolor{red}{b}^\# = \bot$ (the least fixpoint). Intuitively, only the first fixpoint makes sense---the second fixpoint implies that there are no possible values for $\textcolor{red}{b}$. This seems false, because the original program, $1 * 5$, clearly evaluates to $5$. 

The issue is that the rewrite $1 * x \Rightarrow x$ causes the e-graph to represent the graph shown in Figure~\ref{fig:conventional_egraphs_and_eqs_ill_formed_graph}. The cycle from $\textcolor{red}{b}$ to itself prevents a concrete execution from occurring, since the nodes cannot be evaluated in any order that evaluates definitions before their uses. Throughout this paper, we refer to graphs that we cannot assign a semantics to as ``ill-formed''. Despite being ill-formed, we can still generate a set of analysis equations for the ill-formed graph based on its structure (shown in Figure~\ref{fig:conventional_egraphs_and_eqs_ill_formed_eqs}). The least fixpoint of this set of equations is $\textcolor{red}{a}^\# = [1,1], \textcolor{red}{b}^\# = \bot$---because an e-class analysis combines analyses for the same e-class with the meet operator, the least fixpoint for the ill-formed graph will ``poison'' the least fixpoint of the e-class analysis.

This observation seems to run contrary to prior work, where it is assumed that e-class analysis computes a sound abstraction of represented \emph{terms} in the e-graph \cite{egg, egraphs_plus_ai}. In particular, the represented terms of the e-class $\textcolor{red}{b}$ in Figure~\ref{fig:conventional_egraphs_and_eqs_egraph} are $5$, $1 * 5$, $1 * (1 * 5)$, \ldots, all of which evaluate to $5$. The trick that prior work uses, implicitly, is to compute the \emph{greatest} fixpoint of the e-class analysis equations (or a different (post-)fixpoint whose computation is incrementally sound \cite{egraphs_plus_ai}). This circumvents the ``poisoning'' caused by the ill-formed graph in Figure~\ref{fig:conventional_egraphs_and_eqs_ill_formed_graph}, since the greatest fixpoint of the equations in Figure~\ref{fig:conventional_egraphs_and_eqs_ill_formed_eqs} is $\textcolor{red}{a}^\# = [1,1], \textcolor{red}{b}^\# = [-\infty,\infty]$, and for all intervals $a$, $a \sqcap [-\infty,\infty]=a$.

The e-graph shown in Figure~\ref{fig:conventional_egraphs_and_eqs_egraph} is representative of existing use cases, where cycles are only created by rewrite rules that equate a term with one of its sub-terms (for example, rewriting $1*x$ to $x$). In these use cases, all programs of interest are acyclic, and thus computing a greatest fixpoint of the e-class analysis is acceptable. However, for future use cases, we want to incorporate optimism, as programs can contain loops or recursion. We return to the example program shown in Figure~\ref{fig:ssa-graph-example}. Figure~\ref{fig:ssa_egraphs_and_eqs} shows the DFG for this program with the rewrite $1 * x \Rightarrow x$ applied and the e-class analysis equations for the resulting e-graph\footnote{Unlike the graph in Figure~\ref{fig:conventional_egraphs_and_eqs}, the semantics of a DFG with $\phi$ nodes depends on a walk through a CFG. We will properly define these semantics in Section~\ref{sec:semantics_and_ai_ssa}---for now, this analysis can be understood to be flow-insensitive.}. These equations have three fixpoints:

\begin{enumerate}
    \item $\textcolor{red}{a}^\# = [1,1], \textcolor{red}{b}^\# = \bot, \textcolor{red}{c}^\# = [1,1], \textcolor{red}{d}^\# = \bot$
    \item $\textcolor{red}{a}^\# = [1,1], \textcolor{red}{b}^\# = [5,5], \textcolor{red}{c}^\# = [1,\infty], \textcolor{red}{d}^\# = [6,\infty]$
    \item $\textcolor{red}{a}^\# = [1,1], \textcolor{red}{b}^\# = [5,5], \textcolor{red}{c}^\# = [-\infty,\infty], \textcolor{red}{d}^\# = [-\infty,\infty]$
\end{enumerate}

\begin{figure}
    \centering
    \hfill
    \begin{subfigure}[c]{0.4\textwidth}
        \centering
    \begin{tikzpicture}
        \graph[no placement] { 
            phi1[as=$\phi_v^{\textcolor{red}{c}}$,at={(4,0)}];
            pp[as=$+^{\textcolor{red}{d}}$,at={(5, 0)}];
            1[as=$1^{\textcolor{red}{a}}$,at={(5, 1)}];
            t[as=$*^{\textcolor{red}{b}}$,at={(6,0)}];
            5[as=$5^{\textcolor{red}{b}}$,at={(6,1)}];
            1 -> phi1;
            t -> pp;
            pp ->[bend right = 30] phi1;
            phi1 ->[bend right = 30] pp;
            1 -> t;
            5 -> t;
            5 --[red,bend left=50,dashed,thick] t;
        };
    \end{tikzpicture}
        \caption{DFG of program from Figure~\ref{fig:ssa-graph-example}, now as an e-graph, with the rewrite $1 * x \Rightarrow x$.}
        \label{fig:ssa_egraphs_and_eqs_egraph}
    \end{subfigure}
    \hfill
    \begin{subfigure}[c]{0.45\textwidth}
    \centering
    \begin{align*}
        \textcolor{red}{a}^\# &= [1,1] & \textcolor{red}{b}^\# &= [5,5] \sqcap (\textcolor{red}{a}^\# * \textcolor{red}{b}^\#) \\
        \textcolor{red}{c}^\# &= \textcolor{red}{a}^\# \sqcup \textcolor{red}{d}^\# & \textcolor{red}{d}^\# &= \textcolor{red}{b}^\# + \textcolor{red}{c}^\#
    \end{align*}
    \caption{Analysis equations for the DFG in Figure~\ref{fig:ssa_egraphs_and_eqs_egraph}.}
    \label{fig:ssa_egraphs_and_eqs_eqs}
    \end{subfigure}
    \hfill
    \caption{E-class analysis (flow-insensitive, with the interval lattice) of the program from Figure~\ref{fig:ssa-graph-example}.}
    \label{fig:ssa_egraphs_and_eqs}
\end{figure}

Fixpoint \#1 is the least fixpoint and derives $\textcolor{red}{b}^\# = \textcolor{red}{d}^\# = \bot$, which is unsound for \emph{well-formed} represented graphs. Fixpoint \#3 is the greatest fixpoint, so it is sound, but it does not determine that the loop induction variable is always positive. Fixpoint \#2 suffers from neither deficiency (it is sound and optimistic). For a suitable choice of abstract domain (restrictions described in Definition~\ref{def:ai_ssa}), the primary contribution of this work is an algorithm which computes an analysis that is 1) sound for well-formed represented graphs in an e-graph and 2) optimistic, so programs with loops or recursion can be analyzed more precisely (Section~\ref{sec:ai_over_ssa_egraphs}). Prior work in e-graph extraction (which is a kind of e-class analysis) tackles a similar problem by filtering cycles \cite{tensat}. This requires removing e-nodes from the e-graph, which explicitly removes represented graphs. This may harm precision of both rewriting and analysis. Additionally, one cannot remove an e-node from a singleton e-class (otherwise, the e-class would have no e-node to produce during extraction). Our method does not suffer from these issues because it does not modify the e-graph.

\section{Semantics and Abstract Interpretation of SSA Programs}
\label{sec:semantics_and_ai_ssa}

We describe a graph-based SSA form representation that is similar to the sea-of-nodes IR \cite{sea_of_nodes}, but separates data flow and control flow into two graphs (the DFG and the CFG). This separation 1) simplifies specifying the semantics and 2) eases rewriting, as the DFG is easy to treat as an e-graph while the CFG is not. We describe a semantics for this representation, consisting of a denotational component for interpreting the DFG and an operational component for interpreting the CFG. These semantics are similar to prior work describing semantics for the sea-of-nodes \cite{semantic_sea_of_nodes}, with the key difference that the denotational component is parameterized by a particular walk through the CFG, rather than relying on a supplied prior evaluation for $\phi$ nodes---this is used to justify the correctness of our optimistic e-class analysis algorithm in Section~\ref{sec:optimistic_e_class_analysis}. We describe how abstract interpretation can be performed on this representation, including flow-insensitive and flow-sensitive variants.

\subsection{Structure of SSA Programs}
\label{sec:ssa_structure}

A SSA program is a pair $(\mathcal{S}, \mathcal{G})$ of a DFG and a CFG. A DFG $\mathcal{S}$ is a cyclic term (Definition~\ref{def:background_cyclic_terms}) that may contain nodes with $\phi$ function symbols. A CFG $\mathcal{G} = (V, s, P)$ is a graph with vertices $V$, a distinguished entry $s \in V$, and a predecessor function $P \in V \rightarrow (V \times \mathcal{S})^{*}$ which maps each vertex to a tuple of its predecessors---a predecessor is a vertex paired with a node from $\mathcal{S}$, which guards control flow on the edge. We use guard conditions rather than branch instructions, as it makes specifying the semantics simpler. The DFG and CFG are mutually dependent---the CFG uses DFG values as guards, and $\phi$ DFG nodes depend on control flow to select inputs (specifically, $\phi$ function symbols are parameterized by a non-entry vertex from the CFG). For simplicity, we assume that vertices have at most two predecessors, that predecessor vertices are unique, and that all vertices $v$ where the $\phi_v$ function symbol exists in the DFG have exactly two predecessors\footnote{Any program with an arbitrary number of predecessors per CFG vertex can be transformed into an equivalent program with at most two predecessors per vertex. We assume this both in our formalization and in our implementation.}.

\subsection{Concrete Semantics}
\label{sec:ssa_concrete_semantics}

The semantics of SSA programs consists of a denotational component for defining the value of each DFG node (in some concrete domain $\Sigma$, which we assume is a complete lattice) parameterized by a walk in the CFG, and an operational component for defining what walks are possible. We assume that every non-$\phi$ function symbol, $f$, appearing in the DFG can be interpreted as a function $f \in \Sigma^k \rightarrow \Sigma$, where $k$ is the arity of $f$. We also assume that there is a $\mathcal{T} \subseteq \Sigma$ which is the set of ``true'' values (used to evaluate guards). A walk $W$ in a program with DFG $\mathcal{S}$ and CFG $\mathcal{G}=(V,s,P)$ of length $k \in \mathbb{Z}_{>0}$ is a tuple of vertices $W \in V^k$ such that $W_1 = s$ and $\forall i \in \mathbb{Z}, 1 \le i < k \implies (\exists g \in \mathcal{S}, (v_i, g) \in P(v_{i+1}))$---in other words, a walk is a sequence of vertices in $\mathcal{G}$ that starts at $s$ and traverses edges in $\mathcal{G}$. $\mathcal{W}$ is the set of all walks in a program. We often discuss the ``predecessor'' walk of a walk $W$, written as $W'$. We say that $W = W'; v$ when $W$ is the walk $W'$ with the addition of the vertex $v$ at the end.

We first define the denotational semantics of nodes in a DFG $\mathcal{S}$. The denotation of a node is a function from CFG walks to sets of domain values---that is, $\denote{\cdot} \in \mathcal{S} \rightarrow (\mathcal{W} \rightarrow \Sigma)$.

\begin{definition}[Denotation of DFG Nodes]
\label{def:ssa_denotation}
The denotation of a node $n \in \mathcal{S}$ is defined as:
$$
\denote{n}(W) \triangleq \begin{cases}
    \bot 
     & n \ \text{is a} \ \phi_{v} \wedge W = (s) \\
    \denote{n}(W') 
     & n \ \text{is a} \ \phi_{v'} \wedge W = W';v \wedge v \ne v' \\
    \denote{n_1}(W') 
     & n \ \text{is a} \ \phi_{v} \wedge W = W'';p;v \wedge \exists g \in \mathcal{S}, (p, g) = P(v)_1 \\
    \denote{n_2}(W') 
     & n \ \text{is a} \ \phi_{v} \wedge W = W'';p;v \wedge \exists g \in \mathcal{S}, (p, g) = P(v)_2 \\
    f(\denote{n_1}(W), \ldots, \denote{n_k}(W))
     & n \ \text{is a} \ f \in \Sigma^k \rightarrow \Sigma
\end{cases}
$$
\end{definition}

Note that this function is only well defined for certain SSA programs. According to the structure described in Section~\ref{sec:ssa_structure}, a DFG is just a cyclic term---consider the cyclic term shown in Figure~\ref{fig:conventional_egraphs_and_eqs_ill_formed_graph}. $\denote{\textcolor{red}{b}}$ is circularly defined (for any walk in any CFG), and is thus not well defined. To address this issue, we define a subset of SSA programs as \emph{well-formed}.

\begin{definition}[Well-formed SSA Programs]
\label{def:well_formed_ssa}
A SSA program with DFG $\mathcal{S}$ and CFG $\mathcal{G} = (V,s,P)$ is well-formed if and only if:
\begin{enumerate}
    \item Every cycle in $\mathcal{S}$ contains at least one node whose function symbol is a $\phi$ symbol, and\ldots
    \item For every node $n \in \mathcal{S}$ where n is a $\phi_v$, let $((p_1, g_1), (p_2, g_2)) = P(v)$, then,
    \begin{enumerate}
        \item $\forall W \in \mathcal{W}, p_1 \in W \implies \Phi_\text{pred}(n_1) \subseteq W \wedge \Phi_\text{pred}(g_1) \subseteq W$, and\ldots
        \item $\forall W \in \mathcal{W}, p_2 \in W \implies \Phi_\text{pred}(n_2) \subseteq W \wedge \Phi_\text{pred}(g_2) \subseteq W$,
    \end{enumerate}
    where:
    \begin{gather*}
        \Phi_\text{pred}(n) \triangleq \begin{cases}
            \{v\} & n \ \text{is a}\ \phi_v \\
            \bigcup_i \Phi_\text{pred}(n_i) & \text{otherwise}
        \end{cases}
    \end{gather*}
\end{enumerate}
\end{definition}

The first condition enforces that data operations can be evaluated in some order, given values for input $\phi$s. The second condition is the classic strictness condition of SSA form \cite{ssa_book}, which states that every value is defined before it is used along every control flow walk\footnote{More precisely, for every $\phi_v$ node at some vertex $v$, the $\phi$ nodes that are immediately depended on should be defined on all walks that reach $v$---this is true when the vertices of the depended $\phi$ nodes dominate the predecessors of $v$.}. Given a well-formed SSA program and an arbitrary control flow walk, the semantics of any node is well defined.

\newcommand{\walkarrow}{\ensuremath{\rightarrow_{\text{\tiny WALK}}}\xspace}
\newcommand{\walkarrowstar}{\ensuremath{\rightarrow_{\text{\tiny WALK}}^{*}}\xspace}

Next, we define the operational semantics that characterize control flow in a DFG $\mathcal{S}$ and CFG $\mathcal{G} = (V, s, P)$. The core set that is defined is $\mathcal{W}_P$, which is the set of \emph{possible} control flow walks.

\begin{definition}[\walkarrow]
Given $W;v \in \mathcal{W}$ and $v' \in V$, then $W;v \walkarrow W;v;v'$ if and only if $\exists g \in \mathcal{S}, (v, g) \in P(v') \wedge \denote{g}(W;v) \in \mathcal{T}$.
\end{definition}

\begin{definition}[\walkarrowstar and $\mathcal{W}_P$]
$\walkarrowstar$ is the reflexive and transitive closure of $\walkarrow$. The set of possible walks is $\mathcal{W}_P = \{W \in \mathcal{W} | (s) \walkarrowstar W\}$.
\end{definition}

$\walkarrow$ evaluates guards on control flow edges to determine if it is possible to step from a vertex to a successor vertex. Starting from the entry point, all intermediate walks are possible walks.

\subsection{Abstraction of Concrete Semantics}
\label{sec:ai_ssa}

Next, we describe how SSA programs can be abstractly interpreted. For simplicity, we stick to non-relational abstract domains in this paper, as e-class analysis has mostly been limited to non-relational domains in prior work\footnote{\texttt{egglog} supports associating a lattice fact with a tuple of e-classes \cite{egglog}, though we are not aware of an application of \texttt{egglog} that uses this capability to perform relational abstract interpretation. In \cite{egraphs_plus_ai}, there is some discussion of ``relational domains'', but this refers to relational information revealed by rewrites, rather than a relational e-class analysis.}. An abstraction of the concrete semantics of a SSA program will over-approximate the values nodes may evaluate to.

\begin{definition}[Abstract Interpretation of SSA Programs]
\label{def:ai_ssa}
    An abstract interpretation over SSA programs is a pair $(\Sigma^\#, \gamma)$, where:
    \begin{itemize}
        \item $\Sigma^\#$ is a complete lattice\footnote{To distinguish from $\Sigma$, we sometimes write the lattice operations of $\Sigma^\#$ as $\bot^\#$, $\top^\#$, $\sqsubseteq^\#$, $\sqcup^\#$, and $\sqcap^\#$.} with a widening operation $\nabla$ where $\forall s^\#_1, s^\#_2 \in \Sigma^\#, s^\#_1 \sqcup s^\#_2 \sqsubseteq s^\#_1 \nabla s^\#_2$.
        \item $\gamma \in \Sigma^\# \rightarrow (\mathcal{W}_P \rightarrow \Sigma)$ is the concretization function, and is also monotone ($\forall s^\#_1, s^\#_2 \in \Sigma^\#, s^\#_1 \sqsubseteq s^\#_2 \implies \gamma(s^\#_1) \sqsubseteq \gamma(s^\#_2)$) and $\gamma(\top^\#)=\top$.
        \item $\Sigma^\#$ either has only finite ascending chains or all chains $s^\#_0 \in \Sigma^\#$, $s^\#_{j}\triangleq s^\#_{j-1} \nabla t^\#_j$ for arbitrary $t^\#_j \in \Sigma^\#$ are eventually non-increasing (there exists $k \in \mathbb{N}$ such that $s^\#_k \sqsubseteq s^\#_{k-1}$).
        \item For the purpose of computing greatest fixpoints, we assume that all descending chains in $\Sigma^\#$ are finite. This simplifies our definitions and soundness proof in Section~\ref{sec:optimistic_e_class_analysis}. In practice, narrowing can be used to cut off descending chains while maintaining soundness.
    \end{itemize}
\end{definition}

The last two points are necessary to ensure termination. The former is a standard assumption in abstract interpretation \cite{abstract_interpretation}. The latter is a slight modification of a standard assumption in e-class analysis \cite{egraphs_plus_ai}. Additionally, we assume that every non-$\phi$ function symbol $f \in \Sigma^k \rightarrow \Sigma$ has a corresponding abstract transformer $f^\# \in {\Sigma^\#}^k \rightarrow \Sigma^\#$. An abstract interpretation is \emph{sound} when:

\begin{itemize}
    \item $\forall (s^\#_1,\ldots,s^\#_k) \in {\Sigma^\#}^k, f(\gamma(s^\#_1),\ldots, \gamma(s^\#_k)) \sqsubseteq \gamma(f^\#(s^\#_1,\ldots,s^\#_k))$ for all non-$\phi$ function symbols (abstract transformers over-approximate the semantics of concrete functions).
    \item $\forall S^\# \subseteq \mathcal{P}(\Sigma^\#), {\bigsqcup}_{s^\# \in S^\#}\gamma(s^\#) \sqsubseteq \gamma(\bigsqcup S^\#) \: \wedge \: {\bigsqcap}_{s^\# \in S^\#}\gamma(s^\#) \sqsubseteq \gamma(\bigsqcap S^\#)$ (join and meet in the abstract domain over-approximate join and meet in the concrete domain).
    \item $\forall s^\#_1, s^\#_2 \in \Sigma^\#, \gamma(s^\#_1) \sqcup \gamma(s^\#_2) \sqsubseteq \gamma(s^\#_1 \nabla s^\#_2)$ (widening in the abstract domain over-approximates join in the concrete domain).
\end{itemize}

We point out two subtleties. First, $\Sigma^\#$ abstracts $\mathcal{W}_P \rightarrow \Sigma$, not $\Sigma$ (expanded on in Section~\ref{sec:flow-sensitivity}). Second, the co-domain of $\gamma$ is not the same as the co-domain of $\denote{\cdot}$ ($\mathcal{W}_P \rightarrow \Sigma$ vs. $\mathcal{W} \rightarrow \Sigma$). This reflects that $\denote{\cdot}$ is defined over all walks, while we are only interested in abstracting over possible walks. An abstraction $s^\#$ of a node $n$ is \emph{locally} sound for some $W \in \mathcal{W}_P$ if $\denote{n}(W) \sqsubseteq \gamma(s^\#)(W)$. An abstraction $s^\#$ of a node $n$ is sound if and only if it is locally sound for all possible walks.

We can compute a flow-insensitive abstract interpretation on a program by computing a fixpoint of a set of equations---this set of equations contains one equation per node in the DFG. 

\begin{definition}[Equations for Abstract Interpretation of SSA Programs]
\label{def:ssa_eqs}    
Fix a program with DFG $\mathcal{S}$ and CFG $\mathcal{G}=(V,s,P)$. For a node $n \in \mathcal{S}$, its abstraction $n^\#$ can be written:
\begin{itemize}
    \item If $n$ is a $\phi_v$ node for some $v \in V$, then the equation is $n^\# = n_0^\# \sqcup n_1^\#$.
    \item If $n$ is a $f \in \Sigma^k \rightarrow \Sigma$ node, then the equation is $n^\# = f^\#(n^\#_1,\ldots,n^\#_k)$.
\end{itemize}
\end{definition}

A sound abstraction can be computed by 1) starting from $\top$ and iteratively applying the equations, possibly stopping early (narrowing), or 2) starting from $\bot$ and iteratively applying the equations until a (post-)fixpoint is reached, where dependency cycles among the equations are broken with widening (the equation for a widened $\phi$ node $n \in \mathcal{S}$ is $n^\#=n^\#\nabla(n^\#_0\sqcup n^\#_1)$). The $\phi$ nodes whose equations are modified are called ``widening points'', and can be determined as follows: compute a weak topological order (WTO) over the CFG \cite{chaotic_iteration} and mark all $\phi$ nodes whose vertex is a ``component head'' in the WTO as widening points. The WTO identifies a subset of vertices such that all CFG cycles contain at least one node in this subset (called the component heads). All data flow cycles in a well-formed DFG go through $\phi$ nodes, and by strictness, those cycles always correspond to cycles in the CFG. Thus, every cycle in the DFG contains a $\phi$ node that is marked as a widening point.

\subsection{Flow Sensitivity}
\label{sec:flow-sensitivity}

In some program representations, non-relational abstractions are computed at and associated with specific program points. This simplifies computing a flow sensitive analysis, as different abstractions can be stored at different program points. However, in our SSA program representation, every value in the DFG is \emph{global} to the entire function, and the fact that a node may evaluate differently on different walks is lifted into the semantics (this is why the co-domain of $\denote{\cdot}$ is $\mathcal{W} \rightarrow \Sigma$, rather than just $\Sigma$). Thus, abstract objects do not over-approximate concrete values, but rather concrete values per walk. There are \emph{at least} three ways we can build the lattice of abstract objects $\Sigma^\#$:

\begin{enumerate}
    \item $\Sigma^\#$ contains abstractions of concrete values and $\gamma$ returns a constant function ignoring its walk argument. This is often called a \emph{flow insensitive} analysis.
    \item $\Sigma^\#$ contains maps from control flow vertices to abstractions of concrete values and $\gamma$ returns a function that takes the \emph{last} vertex in the passed walk and uses that vertex to index the map being concretized. This is often called a \emph{flow sensitive} analysis.
    \item $\Sigma^\#$ contains maps from (up-to) $k$-tuples of vertices to abstractions of concrete values and $\gamma$ returns a function that takes the \emph{last} (up-to) $k$ vertices in the passed walk and uses that tuple to index the map being concretized. This is sometimes called a $k$-\emph{path sensitive} analysis.
\end{enumerate}

Note that none of these options store an abstraction of concrete values per possible walk\footnote{This is the required lattice to represent a meet-over-paths analysis, which is in general incomputable \cite{monotone_frameworks}.}, since $\mathcal{W}_\mathcal{P}$ may be infinite. In the rest of this paper, we will consider flow insensitive analyses---this is because all prior work in e-class analysis that we are aware of describes flow insensitive analyses \cite{egg, egraphs_plus_ai} and due to a limitation of e-graphs and equality saturation which we expand on in Section~\ref{sec:colors}. We emphasize that there is no fundamental reason that the analysis algorithm we propose in Section~\ref{sec:ai_over_ssa_egraphs} cannot be used for flow sensitive analyses.

\section{Optimistic E-Class Analysis}
\label{sec:ai_over_ssa_egraphs}

The DFG in a SSA program is normally a single graph---this can be replaced by an e-graph which represents multiple graphs. As shown in Section~\ref{sec:challenges}, this has unfortunate consequences when the e-graph represents ill-formed graphs. We expand more on this challenge and propose an abstract interpretation algorithm that computes a sound abstraction of all \emph{well-formed} represented graphs of the e-graph in a SSA program. We also justify why this is the right goal.

\subsection{Representing DFGs with E-Graphs}

Recall that a SSA program consists of a DFG $\mathcal{S}$ and a CFG $\mathcal{G}$. We now replace $\mathcal{S}$ with an e-graph $(\mathcal{N}, \mathcal{C})$. Values in the program are no longer identified by specific nodes, but now by equivalence classes of nodes. Unlike a normal DFG, we do not give direct semantics to an e-graph---this is because the e-graph may represent ill-formed graphs that prevent $\denote{\cdot}$ from being well-defined. 

\begin{definition}[E-Graph Soundness]
\label{def:egraph_soundness}
Fix an e-graph $(\mathcal{N}, \mathcal{C})$ and CFG $\mathcal{G}$. Call $\mathbb{V}$ the set of represented graphs of the e-graph that are also well-formed DFGs, with respect to $\mathcal{G}$---for graph $\mathcal{V} \in \mathbb{V}$, call its representation map $m_\mathcal{V} \in \mathcal{V} \rightarrow \mathcal{N}$. We say that the e-graph is sound if and only if:
\begin{enumerate}
    \item There is a single $\mathcal{W}_P$ such that for all $\mathcal{V} \in \mathbb{V}$, the SSA program consisting of DFG $\mathcal{V}$ and CFG $\mathcal{G}$\footnote{Technically, when combined with an e-graph, $\mathcal{G}$ must be modified to use e-classes as guards rather than nodes in a cyclic term. Then, when considering the program generated by considering a particular represented graph, $\mathcal{G}$ must have each guard e-class substituted with a value from the represented graph that maps into that e-class.} has possible walks $\mathcal{W}_P$, and\ldots
    \item For that $\mathcal{W}_P$, $\forall \mathcal{V}_1,\mathcal{V}_2 \in \mathbb{V}, \forall v_1 \in \mathcal{V}_1, v_2 \in \mathcal{V}_2, [m_{\mathcal{V}_1}(v_1)] = [m_{\mathcal{V}_2}(v_2)] \implies \forall W \in \mathcal{W}_P, \denote{v_1}(W) = \denote{v_2}(W)$.
\end{enumerate}
In other words, all well-formed represented graphs must agree on the set of possible walks (this defines $\mathcal{W}_P$ for a sound e-graph and accompanying CFG) and all nodes in well-formed represented graphs that map into the same e-class must evaluate to the same value on all possible walks.
\end{definition}

Our definition of soundness only concerns equivalence of semantics between nodes of \emph{well-formed} represented graphs. This requirement comes from the use of the semantics of nodes in the represented graphs. As discussed in Section~\ref{sec:ssa_concrete_semantics}, the semantics of SSA programs are only well-defined for well-formed programs. It does not make sense to define e-graph soundness based on represented graphs that do not have well-defined semantics (the ill-formed graphs). In the rest of this paper, we will assume that e-graphs are always sound---an e-graph with singleton e-classes is trivially sound if the initial graph is well-formed and we assume that rewrites preserve soundness.

\subsection{E-Class Analysis of SSA Programs}

E-class analysis associates an abstraction with each e-class in an e-graph. As e-graphs do not have a direct semantics, we use the semantics of well-formed represented graphs to define soundness.

\begin{definition}[E-Class Analysis Soundness]
\label{def:e_class_analysis_soundness}
Fix a sound e-graph $(\mathcal{N}, \mathcal{C})$ and CFG $\mathcal{G}$. Call $\mathbb{V}$ the set of represented graphs of the e-graph that are also well-formed DFGs, with respect to $\mathcal{G}$---for graph $\mathcal{V} \in \mathbb{V}$, call its representation map $m_\mathcal{V} \in \mathcal{V} \rightarrow \mathcal{N}$. $s^\# \in \Sigma^\#$ is a locally sound abstraction for some $W \in \mathcal{W}_P$ of an e-class $c \in \mathcal{C}$ if and only if $\forall \mathcal{V} \in \mathbb{V}, \forall n \in \mathcal{V}, [m_\mathcal{V}(n)] = c \implies \denote{n}(W) \sqsubseteq \gamma(s^\#)(W)$. In other words, all nodes in well-formed represented graphs that map into the e-class $c$ must be over-approximated by $s^\#$. $s^\#$ is a sound abstraction of $c$ if it is locally sound for all $W \in \mathcal{W}_P$.
\end{definition}

As in the definition of soundness of e-graphs, the soundness of an e-class analysis depends only on well-formed represented graphs. Ill-formed represented graphs do not necessarily have a well-defined semantics, so it does not make sense to discuss over-approximating their semantics.

A key property of e-class analysis that is well understood is that sound abstractions of e-nodes in the same e-class can be combined using the lattice meet operator \cite{egg, egraphs_plus_ai}.

\begin{theorem}
\label{theorem:e_class_analysis_meet}
    Fix a sound e-graph $(\mathcal{N}, \mathcal{C})$ and CFG $\mathcal{G}$ with well-formed represented graphs $\mathbb{V}$. For all $\mathcal{V}_1, \mathcal{V}_2 \in \mathbb{V}$ with representation maps $m_{1} \in \mathcal{V}_1 \rightarrow \mathcal{N}, m_{2} \in \mathcal{V}_2 \rightarrow \mathcal{N}$, and for all $n_1 \in \mathcal{V}_1$ with locally sound abstraction $s^\#_1 \in \Sigma^\#$ and $n_2 \in \mathcal{V}_2$ with locally sound abstraction $s^\#_2 \in \Sigma^\#$ (both for some $W \in \mathcal{W}_P$), if $[m_{1}(n_1)] = [m_{2}(n_2)]$, then $s^\#_1 \sqcap s^\#_2$ is a locally sound abstraction for $W$ of $n_1$ and $n_2$.
\end{theorem}

\begin{proof}
    Since $[m_{1}(n_1)] = [m_{2}(n_2)]$, $\denote{n_1}(W)=\denote{n_2}(W)$ by Definition~\ref{def:egraph_soundness}. By assumption, $\denote{n_1}(W) \sqsubseteq \gamma(s^\#_1)(W)$ and $\denote{n_2}(W) \sqsubseteq \gamma(s^\#_2)(W)$. Since $\denote{n_1}(W) = \denote{n_2}(W)$, $\denote{n_1}(W) \sqsubseteq \gamma(s^\#_2)(W)$. Thus, $\denote{n_1}(W) \sqsubseteq \gamma(s^\#_1)(W) \sqcap \gamma(s^\#_2)(W) \sqsubseteq \gamma(s^\#_1 \sqcap s^\#_2)(W)$, so $s^\#_1\sqcap s^\#_2$ is a locally sound abstraction for $W$ of $n_1$. The same argument applies for $n_2$.
\end{proof}

We can perform e-class analysis by analyzing the e-nodes in each e-class and combining the abstractions with the abstract meet operator to get a single ``summary'' abstraction for the e-class.

As in prior work, we can compute e-class analysis as a (post-)fixpoint of a set of equations \cite{egraphs_plus_ai}.

\begin{definition}[Equations for E-Class Analysis of SSA Programs]
\label{def:e_class_analysis_eqs}
Fix a program with sound e-graph $(\mathcal{N}, \mathcal{C})$. Its e-class analysis equations are given by:
\begin{itemize}
    \item For each e-class $c \in \mathcal{C}$, the equation is $c^\# = \bigsqcap_{n \in [c]^{-1}} n^\#$.
    \item For each e-node $n \in \mathcal{N}$, the equation is the same as described in Definition~\ref{def:ssa_eqs}.
\end{itemize}
In practice, the equations for e-nodes are inlined into the equations for e-classes, as in Figure~\ref{fig:conventional_egraphs_and_eqs_eqs}.
\end{definition}

The reason that e-class analyses can be more precise than standard abstract interpretation on individual represented graphs of the e-graph is because abstract interpretation is inherently \emph{intensional}---that is, the way a program is written can affect the computed analysis result, even if the extensional behavior of the program does not change \cite{best_ai, abstract_extensionality}. Performing rewriting in an e-graph with equality saturation exposes equivalences between multiple different ways of writing the same program. These syntactically different but semantically equivalent programs may be analyzed differently and e-class analysis allows us to combine these results to increase precision. We describe a specific example of this phenomenon in Section~\ref{sec:background_eclass_analysis}.

Unfortunately, some (post-)fixpoints of the standard equations for e-class analysis, as given in Definition~\ref{def:e_class_analysis_eqs}, will compute \emph{unsound} e-class analyses for some programs. This was demonstrated with particular examples in Section~\ref{sec:challenges}. The core issue is that while a sound e-class analysis is an over-approximation of the semantics of \emph{well-formed} represented graphs, whether a represented graph is well-formed or ill-formed is a subtle property that is not immediately evident from the structure of the graph---the equations given in Definition~\ref{def:e_class_analysis_eqs} are produced by the immediate structure of an e-graph, and will thus ``consider'' \emph{all} represented graphs. Since e-class analysis combines analysis results with the meet operator, (post-)fixpoints of the resulting equations will not necessarily over-approximate the semantics of well-formed represented graphs of the e-graph.

\subsection{Computing a Sound and Optimistic E-Class Analysis}
\label{sec:optimistic_e_class_analysis}

We propose an algorithm that computes an e-class analysis of a SSA program which is 1) sound (according to Definition~\ref{def:e_class_analysis_soundness}) and 2) optimistic (is able to analyze loops with some precision). Our algorithm relies on two observations regarding e-graphs and their represented graphs:

\begin{enumerate}
    \item All cycles in an e-graph are either 1) cycles containing $\phi$ nodes that correspond to control flow loops or 2) cycles created by rewrites that equate a term to one of its sub-terms. We will call the first kind of cycle ``well-formed'' and the second kind ``ill-formed''.
    \item If the initial SSA program was well-formed, then after rewriting the set of ill-formed represented graphs is exactly the set of represented graphs containing ill-formed cycles.
\end{enumerate}

Recall from Section~\ref{sec:challenges} that a greatest fixpoint solution to the e-class analysis equations is actually a sound e-class analysis because no optimistic assumptions are made (and thus the computation is incrementally sound). Intuitively, our algorithm will allow optimistic assumptions for well-formed cycles while not allowing optimistic assumptions for ill-formed cycles.

Our algorithm proceeds as follows. First, identify the well-formed cycles in the e-graph. We accomplish this in the same way that widening points are determined. We compute a WTO over the CFG, identify component head vertices, and mark all control flow edges to component heads from a vertex in the same component as ``back'' edges. Well-formed cycles in the e-graph contain $\phi$ nodes whose vertex is a component head, and thus have an input corresponding to a back edge.

Second, compute an e-class analysis that is locally sound for all walks that \emph{do not traverse any back edges}. For any $\phi$ with an input corresponding to a back edge, that input to the $\phi$ can effectively be ignored, since no walk being considered can possibly walk that edge. This breaks all well-formed cycles in the e-graph. The resulting e-graph can be analyzed using standard e-class analysis---specifically, we compute a greatest fixpoint to the e-class analysis equations, since this result is a (locally) sound e-class analysis. Call this result $S^\#_0$.

Third, compute an e-class analysis that is locally sound for all walks that \emph{traverse back edges at most $j$ times}, given a locally sound analysis for all walks that traverse back edges at most $j-1$ times. For any $\phi$ with an input corresponding to a back edge, that input to the $\phi$ can directly look up an analysis result from $S^\#_{j-1}$, rather than the result for that e-class currently being computed. This breaks all well-formed cycles in the e-graph, so we can compute a greatest fixpoint to these e-class analysis equations as well. Call this result $S^\#_{j}$.

Fourth, repeat the third step $l$ total times to compute $S^\#_{l}$. Once $S^\#_l=S^\#_{l+1}$, return $S^\#_l$ as the result.

\begin{figure}
    \centering
    \hfill
    \begin{subfigure}[c]{0.34\textwidth}
        \centering
    \begin{tikzpicture}
        \graph[no placement] { 
            phi1[as=$\phi_v^{\textcolor{red}{c}}$,at={(4,0)}];
            pp[as=$+^{\textcolor{red}{d}}$,at={(5, 0)}];
            1[as=$1^{\textcolor{red}{a}}$,at={(5, 1)}];
            t[as=$*^{\textcolor{red}{b}}$,at={(6,0)}];
            5[as=$5^{\textcolor{red}{b}}$,at={(6,1)}];
            1 -> phi1;
            t -> pp;
            pp ->[bend right = 30,blue] phi1;
            phi1 ->[bend right = 30] pp;
            1 -> t;
            5 -> t;
            5 --[red,bend left=50,dashed,thick] t;
        };
    \end{tikzpicture}
        \caption{E-graph from Figure~\ref{fig:ssa_egraphs_and_eqs_egraph}. The blue edge corresponds to a CFG back edge.}
        \label{fig:dataflow-egraph-example-egraph}
    \end{subfigure}
    \hfill
    \begin{subfigure}[c]{0.6\textwidth}
        \centering
        \begin{tabular}{|c||c|c|c|c|c|}
            \hline
            E-class & $S^\#_{-1}$ & $S^\#_0$ & $S^\#_1$ & $S^\#_2$ \\
            \hline
            \hline
            $a$ & $\bot$ & $[1, 1]$ & $[1,1]$    & $[1,1]$ \\
            \hline
            $b$ & $\bot$ & $[5, 5]$ & $[5,5]$    & $[5,5]$ \\
            \hline
            $c$ & $\bot$ & $[1, 1]$ & $[1,\infty]$  & $[1,\infty]$ \\
            \hline
            $d$ & $\bot$ & $[6, 6]$ & $[6,\infty]$ & $[6,\infty]$ \\
            \hline
        \end{tabular}
        \caption{Trace of optimistic e-class analysis applied to the e-graph (flow insensitive with the interval lattice).}
        \label{fig:dataflow-egraph-example-trace}
    \end{subfigure}
    \hfill
    \caption{A trace of optimistic e-class analysis applied to the program from Figure~\ref{fig:ssa-graph-example}.
    }
    \label{fig:dataflow-egraph-example}
\end{figure}

\begin{definition}[Optimistic E-Class Analysis]
Given a SSA program with sound e-graph $(\mathcal{N}, \mathcal{C})$ and CFG $\mathcal{G} = (V, s, P)$ and a set of back edges $B \in V \rightharpoonup \{1, 2\}$ (partial function from vertex to which incoming edge is a back edge) such that every cycle in $\mathcal{G}$ contains an edge from $B$\ldots
\begin{itemize}
    \item $\mathcal{F}_j \in \mathcal{N} \rightarrow \Sigma^\#$ is the abstract transformer for e-nodes during the $j$th iteration.
    \item $S^\#_j \in \mathcal{C} \rightarrow \Sigma^\#$ is the intermediate e-class analysis result during the $j$th iteration.
    \item $\mathcal{F}_{-1} = [n \in \mathcal{N} \mapsto \bot]$.
    \item $\mathcal{F}_j(n) \triangleq \begin{cases}
        n^\#_1 \sqcup n^\#_2 & n\ \text{is a}\ \phi_v \wedge v \not\in \text{dom}(B) \\
        \mathcal{F}_{j-1}(n)\nabla(S^\#_{j-1}(n_1) \sqcup n^\#_2) & n\ \text{is a}\ \phi_v \wedge B(v) = 1 \\
        \mathcal{F}_{j-1}(n)\nabla(n^\#_1 \sqcup S^\#_{j-1}(n_2)) & n\ \text{is a}\ \phi_v \wedge B(v) = 2 \\
        f^\#(n^\#_1, \ldots, n^\#_k) & n\ \text{is a}\ f \in \Sigma^k \rightarrow \Sigma \\
    \end{cases}$, for $j \ge 0$.
    \item $S^\#_{-1} \triangleq [c \in \mathcal{C} \mapsto \bot]$.
    \item $S^\#_j \triangleq S^\#_{j-1} \sqcup \text{gfp}(\{c^\# = \bigsqcap_{n \in [c]^{-1}} \mathcal{F}_j(n) | c \in \mathcal{C}\})$, for $j \ge 0$\footnote{The join with $S^\#_{j-1}$ ensures that $S^\#$ is non-decreasing if $\nabla$ is not monotone. The join can be dropped if $\nabla$ is monotone.}.
    \item The final optimistic e-class analysis result is $S^\#_l$, where $S^\#_l=S^\#_{l+1}$.
\end{itemize}
\end{definition}


\begin{theorem}
    Given a SSA program with sound e-graph $(\mathcal{N}, \mathcal{C})$ and CFG $\mathcal{G}=(V,s,P)$, back edges $B \in V \rightharpoonup \{1, 2\}$, and some $l$ where $S^\#_l=S^\#_{l+1}$, then $S^\#_l$ is a sound e-class analysis of the program.
\end{theorem}

\begin{proof}
    Fix a well-formed represented graph of the e-graph, called $\mathcal{V}$, and its representation map, $m \in \mathcal{V} \rightarrow \mathcal{N}$. We will show that $S^\#_l$ is a sound abstraction of the SSA program $\mathcal{V}$, $\mathcal{G}$.
    
    First, we show that $S^\#_0$ is a locally sound abstraction for all possible walks that do not traverse an edge in $B$. Fix such a walk $W$. $S^\#_0$ is at least the greatest fixpoint of a set of equations---we proceed by showing that 1) the initial assignments of $c^\#$ for $c \in \mathcal{C}$ are locally sound and 2) for each equation, if the previous values of $c^\#$ for $c \in \mathcal{C}$ are locally sound, then the new value for some $c^\#$ computed by the equation is still locally sound. 
    
    The initial assignments for each e-class is $\top^\#$, since we are computing a greatest fixpoint. $\forall s \in \Sigma, s \sqsubseteq \top$, so $\forall n \in \mathcal{V}, \denote{n}(W) \sqsubseteq \top = \gamma(\top^\#)$.

    Fix a node $n \in \mathcal{V}$, its e-node $n' = m(n)$, and its e-class $c = [n']$. If $n$ is a $f \in \Sigma^k \rightarrow \Sigma$ node with sound abstract transformer $f^\#$, then $\mathcal{F}_0(n') = f^\#(n'^\#_1, \ldots, n'^\#_k)$ is a sound abstraction of $n$. If $n$ is a $\phi_v$ node for some $v \in V$, there are three cases. In the first, neither input to the $\phi$ corresponds to an edge in $B$ ($v \not\in \text{dom}(B)$). Thus, $W$ may traverse either predecessor of $v$. $\mathcal{F}_0(n') = n'^\#_1 \sqcup n'^\#_2$---the abstract lattice join operator over-approximates $\phi$ nodes, so this is an over-approximation of $n$. In the second and third cases, one input to the phi corresponds to an edge in $B$---without loss of generality, let us say $B(v) = 1$. $\mathcal{F}_0(n') = \mathcal{F}_{-1}(n')\nabla(S^\#_{-1}(n'_1) \sqcup n'^\#_2) \sqsupseteq n'^\#_2$. This over-approximates $n$, since $\denote{n}(W)$ is either 1) $\bot$ if $W$ has not traversed $v$ or 2) $\denote{n_2}(W')$ if $W$ has traversed $v$, both of which are over-approximated by $n'^\#_2$. $\mathcal{F}_0(n')$ is a locally sound abstraction of $n$ in all cases---by Theorem~\ref{theorem:e_class_analysis_meet}, $\bigsqcap_{n''\in[c]^{-1}} \mathcal{F}_0(n'')$ is a locally sound abstraction of $n$ as well. This is the equation for the abstraction of the e-class $c$, so the equation preserves local soundness.

    Second, we show that if $S^\#_j$ is a locally sound abstraction for all possible walks that traverse edges from $B$ at most $j$ times, then $S^\#_{j+1}$ is a locally sound abstraction for all possible walks that traverse edges from $B$ at most $j+1$ times. Fix a possible walk $W$ that traverses edges from $B$ at most $j+1$ times. We proceed similarly as in the $S^\#_0$ case. $S^\#_{j+1}$ is at least a particular greatest fixpoint---we have already shown that $[c \in \mathcal{C} \mapsto \top^\#]$ is locally sound.

    Fix a node $n \in \mathcal{V}$, its e-node $n'=m(n)$, and its e-class $c= [n']$. If $n$ is a $f \in \Sigma^k \rightarrow \Sigma$ node with sound abstract transformer $f^\#$, then $\mathcal{F}_{j+1}(n')$ is a locally sound abstraction of $n$. If $n$ is a $\phi_v$ node for some $v \in V$, there are three cases. In the first case, neither input to the $\phi$ corresponds to an edge in $B$. This case is shown in the same way as for $S^\#_0$. In the second and third cases, one input to the phi corresponds to an edge in $B$---without loss of generality, let us say $B(v)=1$. $\mathcal{F}_{j+1}(n') = \mathcal{F}_j(n')\nabla(S^\#_j(n'_1) \sqcup n'^\#_2) \sqsupseteq S^\#_j(n'_1) \sqcup n'^\#_2$. Assume that $W$ ends in $v$ (if it does not, then either $\denote{n}(W)=\denote{n}(W')$, so consider the predecessor walk $W'$ instead, or $W=(s)$ so $\denote{n}(W)=\bot$, which is over-approximated by any abstraction). If this walk traversed the first predecessor of $v$, then the predecessor walk $W'$ traversed edges from $B$ at most $j$ times---this is because the edge from the first predecessor of $v$ to $v$ is a back edge. Thus, $\denote{n}(W) = \denote{n_1}(W')$, for which $S^\#_j(n'_1)$, $S^\#_j(n'_1) \sqcup n'^\#_2$, and $\mathcal{F}_{j+_1}(n')$ are locally sound abstractions. If $W$ traversed the second predecessor of $v$, then $\denote{n}(W) = \denote{n_2}(W')$, for which $n'^\#_2$, $S^\#_j(n'_1) \sqcup n'^\#_2$, and $\mathcal{F}_{j+_1}(n')$ are locally sound abstractions. $\mathcal{F}_{j+1}(n')$ is a locally sound abstraction of $n$ in all cases---by Theorem~\ref{theorem:e_class_analysis_meet}, $\bigsqcap_{n''\in[c]^{-1}} \mathcal{F}_{j+1}(n'')$ is a locally sound abstraction of $n$ as well. This is the equation for the abstraction of the e-class $c$, so the equation preserves local soundness.

    Every walk is finite, so every walk visits back edges a finite number of times. If a possible walk visits back edges at most $l$ times, then $S^\#_l$ is a locally sound e-class analysis. Consider a possible walk that visits back edges at most $l'$ times, for some $l' > l$. Since $S^\#_l=S^\#_{l+1}$, and every $S^\#_j$ is defined directly in terms of $S^\#_{j-1}$, it must be the case that $S^\#_l=S^\#_{l+1}=S^\#_{l+2}=\ldots=S^\#_{l'}$, so $S^\#_l$ is still a locally sound e-class analysis. Thus, $S^\#_l$ is a locally sound e-class analysis for all possible walks.
\end{proof}

\subsection{Precision and Complexity}
\label{sec:precision_and_complexity}

Notably, the proof of correctness given in Section~\ref{sec:optimistic_e_class_analysis} assumes very little of the set of back edges\footnote{The only assumption is that $B$ is a partial function, so there is no $v \in V$ such that $(v, 1) \in B \wedge (v, 2) \in B$.}. However, it is advantageous to pick this set carefully for precision and complexity reasons. If this set is too small, then some loops in the original program will not have an edge represented---these loops will not be analyzed optimistically. If this set is too large, then there will be ``too many'' widening points. It is well known that minimizing the set of points at which widening occurs is advantageous for precision \cite{chaotic_iteration}. Figure~\ref{fig:dataflow-egraph-example} shows an example trace of optimistic e-class analysis being applied to the running example program---the result is more precise than standard e-class analysis.

Given $S^\#_{j}$, the complexity to compute $S^\#_{j+1}$ scales with the number of e-nodes, since this bounds the number of equations. As in prior work on e-class analysis, we use a worklist algorithm that propagates the analysis equations for each e-node one-at-a-time \cite{egg}. The maximum amount of times any single equation can be propagated is the maximum length of a descending chain in the abstract lattice. This length is theoretically large, except 1) since this is an incrementally (locally) sound analysis, the length can be cut off with narrowing and 2) we show in Section~\ref{sec:evaluation} that the average number of times equations must be propagated before a fixpoint is reached is very small. 

The quantity $l$ for which $S^\#_l=S^\#_{l+1}$ is bounded by the maximum length of an ascending chain in the abstract lattice (shortened with widening) times $|\{n \in \mathcal{N} | n\ \text{is a}\ \phi_v \wedge v \in \text{dom}(B)\}|$. This is the same bound as the maximum number of iterations over a connected component of the ``iterative'' chaotic iteration strategy that is commonly used in the abstract interpretation literature \cite{chaotic_iteration}. We also show that in practice $S^\#_l=S^\#_{l+1}$ for a very small $l$ in Section~\ref{sec:evaluation}. 

Call $D$ the maximum length of a descending chain (possibly with narrowing), $A$ the maximum length of an ascending chain (possibly with widening), and $L = |\{n \in \mathcal{N} | n\ \text{is a}\ \phi_v \wedge v \in \text{dom}(B)\}|$. The complexity of optimistic e-class analysis is $O(|\mathcal{N}|*D*A*L)$, though we show in Section~\ref{sec:evaluation} that this is a very conservative bound.

\section{Optimism in Equality Saturation}
\label{sec:ai_plus_eqsat}

We now consider how we can combine optimistic e-class analysis with equality saturation. We propose an algorithm that alternates between phases of optimistic e-class analysis and equality saturation, using the improved precision in one half to improve the precision of the other half.

\subsection{Using Analysis Results in Equality Saturation}
\label{sec:analysis_in_eqsat}

First, we discuss how analysis results can help equality saturation. Recall from Section~\ref{sec:background_eqsat} that equality saturation consists of the repeated application of \emph{rewrite rules} to an e-graph. A rewrite rule performs \emph{e-matching} to find represented terms matching some pattern---for each matched term, an \emph{action} is taken, usually to insert a new represented term and assert that sets of e-classes are now known equal. A rewrite rule may additionally depend on a set of \emph{conditions}---that is, even if the pattern e-matches some represented term, some analysis fact must also be known about involved e-classes for the action to be sound. For example, in a setting involving bitvector operations, the rewrite $a / b \Rightarrow a \gg log_2(b)$ is only valid if $b$ is known to be a power of two \cite{hydra}. 

To depend on $\Sigma^\#$ in rewrite rules, we require that $\Sigma^\#$ is a flow-insensitive abstraction. Rewrite rules equate e-classes \emph{globally} and for the e-graph to be sound, it must be true that all e-nodes in the same e-class evaluate to the same value on all possible walks. Thus, a rewrite rule can only use an abstraction that is sound on all possible walks. We discuss a potential relaxation of this requirement in Section~\ref{sec:colors}. Additionally, we require that if a rewrite rule fires given an e-class analysis $S^\#$, it must fire given an e-class analysis ${S^\#}'$ where ${S^\#}' \sqsubseteq S^\#$. This ensures that if analysis results improve, the set of applicable rewrites only grows.

\subsection{Combined Optimistic E-Class Analysis and Equality Saturation}
\label{sec:combined_algo}

Standard e-class analysis can be run at the same time as rewrites and/or rebuilding, where the definition of saturation is extended to include that abstractions of e-classes stop changing \cite{egg, egglog}. This is only true because standard e-class analysis is incrementally sound. This is not necessarily true in optimistic e-class analysis---for arbitrary $j$, $S^\#_j$ is not necessarily a sound e-class analysis. It is well known that optimistic analyses can be combined via the \emph{reduced product} \cite{survey_product_operators}. Unfortunately, existing versions of equality saturation are pessimistic: 1) it is not clear how a ``$\bot$ e-graph'' could be stored, 2) rewrites are never thrown away and assert soundness as a pre-condition and post-condition, and 3) saturation is not guaranteed, so being able to stop early is essential in practice.

\begin{algorithm}[t]
\small
\caption{Optimistic e-class analysis and equality saturation combination}\label{alg:combining}
\begin{algorithmic}[1]
\Procedure{OptimismInEqualitySaturation}{$(\mathcal{N}, \mathcal{C}), \mathcal{G}$}
\State $B \gets \Call{WTOBackEdges}{\mathcal{G}}$ \hfill // $\mathcal{G}$ does not change during analysis or rewriting
\State $S^\# \gets [c \in \mathcal{C} \mapsto \top]$
\While{$(\mathcal{N}, \mathcal{C})$ has changed and before timeout} \label{code:outer}
\State $S^\# \gets S^\# \sqcap \Call{OptimisticEClassAnalysis}{(\mathcal{N}, \mathcal{C}), B}$ \label{code:reinit}
\While{$(\mathcal{N}, \mathcal{C})$ has changed and before timeout}
\State Apply rewrites to $(\mathcal{N}, \mathcal{C})$ using $S^\#$
\State Perform rebuilding on $(\mathcal{N}, \mathcal{C})$
\EndWhile
\EndWhile \label{code:end_outer}
\State \Return{$(\mathcal{N}, \mathcal{C}), S^\#$}
\EndProcedure
\end{algorithmic}
\end{algorithm}

We propose that optimistic e-class analysis and equality saturation be run in two separate halves---these halves execute repeatedly and results are propagated between the halves. This ensures that only sound results derived from the e-class analysis are used for rewrites. Algorithm~\ref{alg:combining} shows pseudo-code for this approach, which we call ``optimism in equality saturation''.

Due to aforementioned restrictions on rewrite rules, more precise abstractions will never result in an e-graph with fewer rewrites applied. However, an e-graph with more applied rewrites will not necessarily result in at least as precise an analysis, because widening is not necessarily monotone. Since every intermediate e-graph is sound, any intermediate analysis result is also sound---therefore, the intermediate analysis $S^\#$ can be forced to never increase by meeting it with the previous iteration's result (line~\ref{code:reinit}). With this modification, the precision of both the rewriting and analysis halves either improve each iteration or do not change. Every intermediate step of this algorithm stores 1) a sound e-graph and 2) a sound analysis result. Therefore, both loops may be run on a timeout or a maximum number of iterations, as is common in equality saturation applications.

\subsection{Using Flow Sensitive Abstractions in Contextual E-Graphs}
\label{sec:colors}

To use flow sensitive abstractions in rewrites, one would need to be able to record equivalences in a flow sensitive manner. However, the equivalence relation that an e-graph stores is flow insensitive (the equivalences between nodes must hold on \emph{all} possible walks). Conceptually, we can modify the equivalence relation to be flow sensitive. Then, when a rewrite rule depends on a flow sensitive abstraction, any discovered equalities are recorded flow sensitively. In prior work, e-graphs that can store multiple equivalence relations efficiently have been called ``contextual e-graphs'' \cite{contextual_eqsat, relational_contextual_eqsat} or ``colored e-graphs'' \cite{colored_egraphs, easter_egg}. Equivalence relations are identified by either ``contexts'' or ``colors'' and form a hierarchy or lattice relationship. Intuitively, if a context $A$ is the parent of another context $B$ in the hierarchy, then any equalities known in $A$ are automatically known in $B$. For implementing a flow sensitive equivalence relation with a contextual e-graph, we suggest that dominance in a CFG is a good choice of hierarchy. Implementing contextual e-graphs efficiently is an open research question, and we hope this perspective serves as a motivating use case for future work.

\section{Examples and Evaluation}
\label{sec:implementation_and_examples}

In this section, we describe how optimism in equality saturation applies to a set of example programs. We implement optimistic e-class analysis and equality saturation in a small Rust tool that correctly analyses the example programs. We also evaluate this tool on a set of randomly generated programs to 1) confirm that the rewritten e-graphs and e-class analyses are indeed sound and 2) evaluate the performance of optimistic e-class analysis, compared with standard abstract interpretation and standard e-class analysis. All reported wall clock times constitute an average of 25 runs. All experiments were run on a standard laptop (Intel Core Ultra 7 155H @ 1.4 GHz w/  32GB LPDDR5 memory). The code for the implementation, its test suite, and its empirical evaluation are all available in the software artifact accompanying this paper \cite{artifact}.

\subsection{Concrete Implementation}

We implemented optimism in equality saturation for SSA programs in a small Rust program. The tool parses programs in a pseudo-code syntax and translates each function into a SSA program. Optimistic e-class analysis and equality saturation are run as described in Section~\ref{sec:combined_algo}.

We implement three analyses---intervals, global value numbering (GVN), and reachability. Interval analysis is the standard non-relational abstraction of numeric values consisting of a range from a low integer (or $-\infty$) to a high integer (or $\infty$) \cite{abstract_interpretation, systematic_program_analysis}. GVN analysis computes a value number per e-node in the e-graph---a conditional rewrite merges e-nodes whose value numbers are the same. $\bot_\text{GVN}$ is a placeholder that is equal to itself, so back edge inputs to phis are optimistically assumed to be equal---this is what allows us to de-duplicate isomorphic well-formed cycles. $\top_\text{GVN}$ is a placeholder that is considered not equal to itself, so the greatest fixpoint computation during optimistic e-class analysis is incrementally locally sound. Reachability analysis abstracts what control flow walks are possible. We split reachability analysis into two cooperating abstractions---vertex reachability and edge reachability. As reachability analysis characterizes the CFG rather than the SSA e-graph, we implement it as a separate cooperating analysis in a reduced product with the optimistic e-class analysis. The reachability abstraction makes the interval and GVN abstractions more precise by eliminating unreachable edge inputs to $\phi$ nodes. The interval abstraction makes the reachability abstraction more precise by identifying when an edge's condition will never be met (in our implementation, an edge will never be met if the interval $[0, 0]$ can be derived for the condition value)---this effectively captures sparse conditional constant propagation \cite{sccp}.

Notably, interval domains have either long descending chains or infinite descending chains, depending on the exact implementation. This violates an assumption (given in Definition~\ref{def:ai_ssa}) we use for our formal description of optimistic e-class analysis. In a concrete implementation, we can compute a \emph{bounded} sequence of incrementally locally sound abstractions, rather than a greatest fixpoint, to maintain soundness while allowing our algorithm to terminate. This matches observations made in prior work regarding the termination of e-class analysis \cite{egraphs_plus_ai}. In practice, our tool is always able to compute greatest fixpoints for every input program used in our evaluation.

\begin{figure}
    \centering
    \hfill
    \begin{minipage}{0.51\linewidth}
        \begin{lstlisting}[language=Rust,escapechar=|]
fn example1(y) {
    let x = -6;
    let z = 42;
    while y < 10 {
        y = y + 1;
        x = x + 8;
        let lhs = ((x + y) + z) * y; |\label{code:use_x}|
        let rhs = 2 * y + (y * y + z * y);
        if lhs != rhs {|\label{code:program1_if}|
            z = 24;|\label{code:program1_unreach}|
        }
        x = x - 8;
    }
    return z + 7;
}\end{lstlisting}
        \caption{
         First example program to analyze. The goal is to show that the returned value is \texttt{49}, which requires rewriting, reachability analysis, and interval analysis.
        }
        \label{fig:first-code-example}
    \end{minipage}
    \hfill
    \begin{minipage}{0.353\linewidth}
        \begin{lstlisting}[language=Rust,escapechar=|]
fn example2(x) {
    let y = x;
    while y < 10 {
        let xt = x;
        x = y * y + y * 5;
        y = xt * (y + 5 + 0);
    }
    return x - y;|\label{code:program2_return}|
}\end{lstlisting}
        \caption{Second example program to analyze. The goal is to show that the returned value is \texttt{0}, which requires rewriting and GVN analysis.}
        \label{fig:second-code-example}
    \end{minipage}
    \hfill
\end{figure}

\subsection{Example Programs}

The first example program is shown in Figure~\ref{fig:first-code-example}. The ``goal'' is to discover that the return value is equal to 49. At a high level, discovering this fact requires an optimistic interval analysis to discover that $\texttt{x} = 2$ at line~\ref{code:use_x}\footnote{We define flow insensitive abstractions over \emph{SSA values}, not \emph{program variables}, so our tool can find abstractions of \texttt{x} at different locations when \texttt{x} corresponds to different values at those locations.}, rewrites to discover that $\texttt{((2 + y) + z) * y} = \texttt{2 * y + (y * y + z * y)}$, a rewrite to discover that $(\texttt{2 * y + (y * y + z * y) != 2 * y + (y * y + z * y)}) = 0$, and finally a reachability analysis to discover that line~\ref{code:program1_unreach} is unreachable.

The second example program is shown in Figure~\ref{fig:second-code-example}. The ``goal'' is to discover that the return value is equal to 0. Since \texttt{x} is a function parameter, its interval is $[-\infty, \infty]$---to discover $\texttt{x} = \texttt{y}$, the GVN analysis must find that the SSA values being subtracted at the end of the function have the same value number. Rewriting discovers that $\texttt{xt * (y + 5 + 0)} = \texttt{xt * y + xt * 5}$. Then, the optimistic GVN analysis can derive that $\texttt{x}$ and $\texttt{y}$ share a value number, and a rewrite merges their e-classes. The final return constant value is discovered by the rewrite $x - x\Rightarrow0$. Note that we include the rewrite rule $x + 0 \Rightarrow x$ in our rule set, so we discover an ill-formed cycle due to the sub-expression \texttt{5 + 0}. This ill-formed cycle does not poison our optimistic e-class analysis.

The second example demonstrates a capability of optimistic e-class analysis that has eluded prior equality saturation systems: soundly de-duplicating cycles. Our rewrite set includes a rule that merges the e-classes of two e-nodes if the GVN analysis derives the same value number for both e-nodes. Since the GVN analysis is optimistic, two well-formed cycles can be discovered to share value numbers (more precisely, there is a mapping between e-classes in each cycle such that each pair of e-classes in the mapping share value numbers). In effect, this allows us to merge well-formed cycles, which is sound, while not merging ill-formed cycles, which is unsound. This addresses a known problem in the e-graphs literature \cite{e_peg, co_egraphs, omelets_need_onions}.

We wrote these examples both in a pseudo-code language for our tool (shown in Figures~\ref{fig:first-code-example}~and~\ref{fig:second-code-example}) and in C---we compiled the examples in C with GCC 15.2 and Clang 21.1.0 (with optimization level \texttt{-O2}). Neither GCC nor Clang can fully optimize either example. We ran our tool on both examples in three configurations: 1) standard abstract interpretation before rewriting, 2) standard e-class analysis during rewriting, and 3) optimistic e-class analysis during rewriting. Neither standard abstract interpretation nor standard e-class analysis derives the correct constant for the return value of each function, while optimism in equality saturation does.

\subsection{Evaluation on Randomly Generated Programs}
\label{sec:evaluation}

We randomly generated 100 programs of varying sizes to 1) confirm that optimism in equality saturation can soundly analyze and rewrite programs and 2) empirically evaluate the performance of optimistic e-class analysis. After rewriting, the largest e-graph contained 11375 e-nodes and the most loops in a single program was 184.

First, we ran each program in a reference interpreter and checked if optimistic e-class analysis computed over-approximations of the programs' concrete behaviors (before and after rewriting). The analyses were sound for all generated programs (this should not be surprising---we proved that optimistic e-class analysis computes a sound abstraction of well-formed represented graphs).

\begin{table}[t]
    \centering
    \footnotesize
    \begin{tabular}{|c|c|c|c|c|}
         \hline
        & Min & Median & Mean & Max \\
         \hline
        \makecell{Standard Abstract \\ Interpretation (\textmu s)} & 5.87 & 252.95 & 755.35 & 7244.56 \\
         \hline
        \makecell{Standard E-Class \\ Analysis (\textmu s)} & 3.61 & 160.87 & 478.59 & 2320.57 \\
         \hline
        \makecell{Optimistic E-Class \\ Analysis (\textmu s)} & 6.16 (0.95x, 1.57x) & 448.64 (1.44x, 3.01x) & 1480.11 (1.88x, 2.94x) & 7723.35 (34.20x, 5.28x) \\
         \hline
        \makecell{\# Visit Items \\ Pre-rewriting} & 47.00 & 721.00 & 1405.57 & 9271.00 \\
         \hline
        \makecell{\# Visit Items \\ Post-rewriting} & 50.00 (0.98x) & 1222.00 (1.35x) & 3034.69 (1.78x) & 12993.00 (19.75x) \\
         \hline
        \makecell{$n$ for Standard \\ Abstract Interpretation} & 2.00 & 3.00 & 2.95 & 6.00 \\
         \hline
        \makecell{$n$ for Optimistic \\ E-Class Analysis} & 2.00 & 3.00 & 2.95 & 6.00 \\
         \hline
    \end{tabular}
    \caption{Wall clock time (in \textmu s) for standard abstract interpretation, standard e-class analysis, and optimistic e-class analysis run on 100 programs, along with the \# of visit items and the \# of fixpoint computations ($n$). The relative execution time for optimistic e-class analysis is shown in parentheses compared to standard abstract interpretation and standard e-class analysis. We compute the arithmetic mean of absolute measurements and the geometric mean of relative measurements. The \# of visit items after rewriting is shown relative to before. Standard abstract interpretation is run before rewriting while both e-class analyses are run after.}
    \label{tab:evaluation}
\end{table}

Second, we ran optimistic e-class analysis on each program after several rounds of rewriting and measured 1) the number of iterations of the greatest fixpoint computation to compute each $S^\#_j$ and 2) for what $l$ is $S^\#_l=S^\#_{l+1}$. In practice, the greatest fixpoint visits e-nodes, rather than e-classes, and also visits vertices and edges in the CFG (to propagate the reachability analysis). We call each of these objects ``visit items''\footnote{The number of visit items is the number of e-nodes plus the number of CFG vertices plus the number of CFG edges.}. We measured, for each program, the average number of times the greatest fixpoint computation visits each visit item. The average of this average across all generated programs was 3.19 and the maximum of this average was 4.38. In other words, the greatest fixpoint computation visits each visit item a small number of times on average. If for a fixed program we have a minimum $l$ such that $S^\#_l=S^\#_{l+1}$, then optimistic e-class analysis performs $n = l + 2$ greatest fixpoint computations. We measure $n$ for each generated program---the maximum observed $n$ was 6 and for $99\%$ of programs $n$ was 3 or less. Note that the theoretical bound on $n$ is a multiple of the number of $\phi$ e-nodes whose vertex is a component head, which at its largest was 3068.

Third, we measured the execution time of standard abstract interpretation (using iterative chaotic iteration), standard e-class analysis, and optimistic e-class analysis on the generated programs. Our results are summarized in Table~\ref{tab:evaluation}. We observed that in the mean and median cases, the slowdown from using optimistic e-class analysis over standard abstract interpretation was similar to the factor increase in the number of visit items after performing rewriting ($1.44 \approx 1.35$ and $1.88 \approx 1.78$). Additionally, the slowdown over standard e-class analysis was similar to the factor $n$ that corresponds to computing a greatest fixpoint multiple times ($3.01 \approx 3$, $2.94 \approx 2.95$). In other words, the slowdown in using optimistic e-class analysis is associated with, depending on the baseline, either 1) the increase in the size of the e-graph from rewriting or 2) from iteratively computing an abstraction in a similar fashion as the standard iterative chaotic iteration strategy.

\section{Related Work}

\paragraph{E-Graphs and Equality Saturation}

E-graphs were originally developed for computing congruence closure over a set of terms \cite{nelson1980} and are a key ingredient in SMT solvers for propagating equalities between theories \cite{nelson1979}. Later work proposed applying rewrite rules to e-graphs via e-matching during equality saturation \cite{e_peg, eqsat_llvm_tvalid, egg}. 
Most e-matching implementations are based on pattern compilation \cite{efficient_e_matching} or relational queries \cite{relational_e_matching}. 
Another consideration is the representation of the e-graph---traditionally, e-graphs are stored directly as graphs in memory \cite{z3, egg}, while recent work proposes using database tables with a canonicalization procedure \cite{towards_relational_e_graph, egglog}. Acyclic term languages are the most common target for equality saturation. Cycles are often created via rewrite rules, but are rarely desired and often filtered or ignored during extraction \cite{tensat, eboost, smoothe}. The only prior work that we are aware of that encodes an explicitly cyclic program representation into e-graphs is Peggy \cite{e_peg}, which does not support lattice-based analyses\footnote{Some simple analyses, such as constant propagation, are re-implemented as pure syntactic transformations in Peggy.}. Neither \texttt{egg} nor \texttt{egglog}\footnote{\texttt{egglog}'s relational e-matching can theoretically match graph patterns, not just term patterns \cite{relational_e_matching}.} support creating an initial e-graph with cycles and both tools treat extracting cycles as a failure mode \cite{egg, egglog}.

\paragraph{Abstract Interpretation over E-Graphs}

E-class analysis is a technique for performing lattice-based analyses on e-graphs. A fact from a single lattice is assigned to each e-class in an e-graph \cite{egg}. Standard e-class analysis is computed bottom-up and all facts are conservatively initialized to $\top$. Facts are propagated during rewriting, since the analysis is incrementally sound. Later work formalizes e-class analysis as an abstract interpretation and observes the fruitful back-and-forth between analysis and rewriting, but does not produce a precise result for well-formed cycles in a cyclic program representation \cite{egraphs_plus_ai}. \texttt{egglog} allows users to define analyses with functional database tables.
Unlike in \egg, the database approach supports 1) storing multiple analyses per e-class and 2) relational analyses involving multiple e-classes. 
However, the restriction of incremental soundness still applies, which prevents analyzing cyclic programs optimistically. To our knowledge, no prior equality saturation system supports optimistic analyses of well-formed cycles.

\paragraph{Equivalence and Relational Abstractions}

Several relational abstract domains have been proposed, including difference bounds \cite{mine_dbms}, difference abstraction \cite{mine_graph_based}, octagons \cite{mine_octagon, singh_octagon}, pentagons \cite{pentagons}, polyhedra \cite{polyhedra, fast_polyhedra}, Karr's domain \cite{karr}, two variables per inequality \cite{tvpi}, and equalities (using e-graphs to store the equivalence relation) \cite{alien_expressions}. E-class analysis has not yet been extended to arbitrary relational domains---we believe this is an important direction for future work. As relational abstractions store an abstract value per pair, or even tuple, of variables, the entire abstract state can grow quite large. Some prior work takes advantage of sparseness in the relations between variables \cite{singh_octagon}. Another approach uses a data structure called a ``labeled union find'' to store some weakly relational domains by a spanning tree, rather than a fully connected graph \cite{labeled_union_find}. A weakly relational domain can be used to ``map factorize'' compatible non-relational domains, meaning the non-relational domain is only stored per relational class, rather than per program variable. The equivalence relation stored in an e-graph has been viewed by some prior work as an abstraction \cite{alien_expressions} and can be seen as a maximal version of map factorization---e-class analysis stores a fact per e-class, not per e-node, and is compatible with every non-relational abstraction. E-graphs not only factor abstractions by equivalence classes but also \emph{terms}---this is how e-graphs compactly represent a large (sometimes infinite) set of equivalent terms \cite{nelson1980}. An interesting direction for future work could be exploring factoring terms by relations other than equivalence relations---labeled union finds could be used to factor terms that are equivalent modulo a group action, for example \cite{omelets_need_onions}.

\paragraph{Combining Analyses and Transformations}

The structure of many compilers consists of phases of analyses and transformations---either the order of analyses and transformations is designed explicitly \cite{hotspot} or a generic ``pass'' mechanism provides flexible ordering \cite{xla, llvm, mlir}. Every intermediate program is well-formed and optimistic analyses are often employed. However, intermediate programs are ephemeral and transformations are run in a specific order, leading to the classic phase ordering problem. Program transformations have also been used specifically as an ad-hoc mechanism to facilitate communication between program analyses \cite{analyses_plus_transformations}. Another approach to combining analyses and transformations is to view certain transformations themselves as abstract interpretations. Recent work has applied this perspective to SSA translation \cite{ssa_translation_ai} and to simple program simplifications \cite{compiling_with_ai}. Formalizing transformations as abstract interpretations allows for straightforward combinations with standard analyses via product operators \cite{compiling_with_ai}, unlike the combination we propose in Section~\ref{sec:combined_algo}. We believe a version of ``optimistic rewriting'' may be possible to arrive at by formalizing equality saturation as an abstract interpretation. However, we also believe that the implementation of such a method would be difficult to optimize---an efficient implementation of persistent or backtrack-able e-graphs is likely necessary.

\section{Conclusion}

This paper tackles the problem of performing optimistic analyses over e-graphs in tandem with equality saturation. We identify a key problem that can cause optimistic e-class analysis to compute unsound results over e-graphs---equality saturation can create ill-formed represented graphs which poison the analysis with unsoundness. We propose an algorithm to compute optimistic e-class analyses that are sound for all well-formed represented graphs. This technique allows for sound and optimistic analysis of e-graphs containing cyclic program representations during equality saturation. Additionally, optimistic e-class analysis has a similar performance profile as existing abstract interpretation and e-class analysis techniques, meaning it is practical to use.

The treatment in this paper is primarily theoretical and discovers a qualitative improvement in capability over prior equality saturation systems. Important future work includes exploring domains where this technique could uncover quantitative gains in precision. Optimizing SSA programs is certainly a candidate for this kind of investigation. However, we believe that reconsidering the program representation itself could be fruitful---SSA programs separate data flow and control flow, and only data flow is represented in the e-graph. We believe optimism in equality saturation is applicable to other cyclic program representations, such as program expression graphs. Additionally, while we tackle optimistic e-class analysis in this paper, we do not propose an optimistic alternative to equality saturation---such an alternative may possess theoretical and/or practical advantages.

\begin{acks}
We are very grateful for discussions with Tyler Hou, Samuel Coward, Cheng Zhang, Alexandra Silva, George Constantinides, and more broadly at the Dagstuhl Seminar 26022, ``Program Optimization with E-Graphs''. We would also like to thank our reviewers for their valuable feedback. This work is supported by NSF grants IIS-1955488, IIS-2027575, DOE awards DE-SC0016260, AC02-05CH11231, and DARPA Agreement No. HR00112590131.
\end{acks}

\section*{Data Availability Statement}
The tool described in Section~\ref{sec:implementation_and_examples} is open source and freely available at \url{https://github.com/RArbore/pldi26-artifact}. An archived version of the tool along with its evaluation set-up is available as a Docker image hosted on Zenodo \cite{artifact}.

\bibliographystyle{ACM-Reference-Format}
\bibliography{sample-base}

@article{survey_product_operators,
author = {Cortesi, Agostino and Costantini, Giulia and Ferrara, Pietro},
year = {2013},
month = {09},
pages = {},
title = {A Survey on Product Operators in Abstract Interpretation},
volume = {129},
journal = {Electronic Proceedings in Theoretical Computer Science},
doi = {10.4204/EPTCS.129.19}
}

@InProceedings{alien_expressions,
author="Chang, Bor-Yuh Evan
and Leino, K. Rustan M.",
editor="Cousot, Radhia",
title="Abstract Interpretation with Alien Expressions and Heap Structures",
booktitle="Verification, Model Checking, and Abstract Interpretation",
year="2005",
publisher="Springer Berlin Heidelberg",
address="Berlin, Heidelberg",
pages="147--163",
isbn="978-3-540-30579-8"
}

@article{egglog,
author = {Zhang, Yihong and Wang, Yisu Remy and Flatt, Oliver and Cao, David and Zucker, Philip and Rosenthal, Eli and Tatlock, Zachary and Willsey, Max},
title = {Better Together: Unifying Datalog and Equality Saturation},
year = {2023},
issue_date = {June 2023},
publisher = {Association for Computing Machinery},
address = {New York, NY, USA},
volume = {7},
number = {PLDI},
url = {https://doi.org/10.1145/3591239},
doi = {10.1145/3591239},
journal = {Proc. ACM Program. Lang.},
month = jun,
articleno = {125},
numpages = {25}
}

@article{egg,
author = {Willsey, Max and Nandi, Chandrakana and Wang, Yisu Remy and Flatt, Oliver and Tatlock, Zachary and Panchekha, Pavel},
title = {egg: Fast and extensible equality saturation},
year = {2021},
issue_date = {January 2021},
publisher = {Association for Computing Machinery},
address = {New York, NY, USA},
volume = {5},
number = {POPL},
url = {https://doi.org/10.1145/3434304},
doi = {10.1145/3434304},
journal = {Proc. ACM Program. Lang.},
month = jan,
articleno = {23},
numpages = {29}
}

@article{combining_analyses,
author = {Click, Cliff and Cooper, Keith D.},
title = {Combining analyses, combining optimizations},
year = {1995},
issue_date = {March 1995},
publisher = {Association for Computing Machinery},
address = {New York, NY, USA},
volume = {17},
number = {2},
issn = {0164-0925},
url = {https://doi.org/10.1145/201059.201061},
doi = {10.1145/201059.201061},
journal = {ACM Trans. Program. Lang. Syst.},
month = mar,
pages = {181–196},
numpages = {16}
}

@inproceedings{egraphs_plus_ai,
author = {Coward, Samuel and Constantinides, George A. and Drane, Theo},
title = {Combining E-Graphs with Abstract Interpretation},
year = {2023},
isbn = {9798400701702},
publisher = {Association for Computing Machinery},
address = {New York, NY, USA},
url = {https://doi.org/10.1145/3589250.3596144},
doi = {10.1145/3589250.3596144},
booktitle = {Proceedings of the 12th ACM SIGPLAN International Workshop on the State Of the Art in Program Analysis},
pages = {1–7},
numpages = {7},
location = {Orlando, FL, USA},
series = {SOAP 2023}
}

@inproceedings{analyses_plus_transformations,
author = {Lerner, Sorin and Grove, David and Chambers, Craig},
title = {Composing dataflow analyses and transformations},
year = {2002},
isbn = {1581134509},
publisher = {Association for Computing Machinery},
address = {New York, NY, USA},
url = {https://doi.org/10.1145/503272.503298},
doi = {10.1145/503272.503298},
booktitle = {Proceedings of the 29th ACM SIGPLAN-SIGACT Symposium on Principles of Programming Languages},
pages = {270–282},
numpages = {13},
location = {Portland, Oregon},
series = {POPL '02}
}

@inproceedings{contextual_eqsat,
  TITLE = {{Contextual Equality Saturation}},
  AUTHOR = {Drewery, Alexandre and Jensen, Thomas and Pichardie, David},
  URL = {https://inria.hal.science/hal-05226543},
  BOOKTITLE = {{SAS 2025 - 32nd Static Analysis Symposium}},
  ADDRESS = {Singapore, Singapore},
  PAGES = {1-26},
  YEAR = {2025},
  MONTH = Oct,
  HAL_ID = {hal-05226543},
  HAL_VERSION = {v1},
}

@inproceedings{gvn,
author = {Alpern, B. and Wegman, M. N. and Zadeck, F. K.},
title = {Detecting equality of variables in programs},
year = {1988},
isbn = {0897912527},
publisher = {Association for Computing Machinery},
address = {New York, NY, USA},
url = {https://doi.org/10.1145/73560.73561},
doi = {10.1145/73560.73561},
booktitle = {Proceedings of the 15th ACM SIGPLAN-SIGACT Symposium on Principles of Programming Languages},
pages = {1–11},
numpages = {11},
location = {San Diego, California, USA},
series = {POPL '88}
}

@article{sccp,
author = {Wegman, Mark N. and Zadeck, F. Kenneth},
title = {Constant propagation with conditional branches},
year = {1991},
issue_date = {April 1991},
publisher = {Association for Computing Machinery},
address = {New York, NY, USA},
volume = {13},
number = {2},
issn = {0164-0925},
url = {https://doi.org/10.1145/103135.103136},
doi = {10.1145/103135.103136},
journal = {ACM Trans. Program. Lang. Syst.},
month = apr,
pages = {181–210},
numpages = {30}
}

@InProceedings{chaotic_iteration,
author="Bourdoncle, Fran{\c{c}}ois",
editor="Bj{\o}rner, Dines
and Broy, Manfred
and Pottosin, Igor V.",
title="Efficient chaotic iteration strategies with widenings",
booktitle="Formal Methods in Programming and Their Applications",
year="1993",
publisher="Springer Berlin Heidelberg",
address="Berlin, Heidelberg",
pages="128--141",
isbn="978-3-540-48056-3"
}

@article{e_peg,
author = {Tate, Ross and Stepp, Michael and Tatlock, Zachary and Lerner, Sorin},
title = {Equality saturation: a new approach to optimization},
year = {2009},
issue_date = {January 2009},
publisher = {Association for Computing Machinery},
address = {New York, NY, USA},
volume = {44},
number = {1},
issn = {0362-1340},
url = {https://doi.org/10.1145/1594834.1480915},
doi = {10.1145/1594834.1480915},
journal = {SIGPLAN Not.},
month = jan,
pages = {264–276},
numpages = {13}
}

@misc{omelets_need_onions,
      title={Omelets Need Onions: E-graphs Modulo Theories via Bottom-up E-matching}, 
      author={Philip Zucker},
      year={2025},
      eprint={2504.14340},
      archivePrefix={arXiv},
      primaryClass={cs.PL},
      url={https://arxiv.org/abs/2504.14340}, 
}

@article{scc_gvn,
author = {Simpson, Taylor and Cooper, Keith and Simpson, L.},
year = {1997},
month = {02},
pages = {},
title = {SCC-based value numbering}
}

@misc{relational_contextual_eqsat,
      title={Towards Relational Contextual Equality Saturation}, 
      author={Tyler Hou and Shadaj Laddad and Joseph M. Hellerstein},
      year={2025},
      eprint={2507.11897},
      archivePrefix={arXiv},
      primaryClass={cs.PL},
      url={https://arxiv.org/abs/2507.11897}, 
}

@article{ssa_translation_ai,
author = {Lemerre, Matthieu},
title = {SSA Translation Is an Abstract Interpretation},
year = {2023},
issue_date = {January 2023},
publisher = {Association for Computing Machinery},
address = {New York, NY, USA},
volume = {7},
number = {POPL},
url = {https://doi.org/10.1145/3571258},
doi = {10.1145/3571258},
journal = {Proc. ACM Program. Lang.},
month = jan,
articleno = {65},
numpages = {30}
}

@article{compiling_with_ai,
author = {Lesbre, Dorian and Lemerre, Matthieu},
title = {Compiling with Abstract Interpretation},
year = {2024},
issue_date = {June 2024},
publisher = {Association for Computing Machinery},
address = {New York, NY, USA},
volume = {8},
number = {PLDI},
url = {https://doi.org/10.1145/3656392},
doi = {10.1145/3656392},
journal = {Proc. ACM Program. Lang.},
month = jun,
articleno = {162},
numpages = {26}
}

@article{combining_program_improvers,
author = {Veldhuizen, Todd and Siek, Jeremy},
year = {2003},
month = {04},
pages = {},
title = {On Combining Program Improvers}
}

@inproceedings{systematic_program_analysis,
author = {Cousot, Patrick and Cousot, Radhia},
title = {Systematic design of program analysis frameworks},
year = {1979},
isbn = {9781450373579},
publisher = {Association for Computing Machinery},
address = {New York, NY, USA},
url = {https://doi.org/10.1145/567752.567778},
doi = {10.1145/567752.567778},
booktitle = {Proceedings of the 6th ACM SIGACT-SIGPLAN Symposium on Principles of Programming Languages},
pages = {269–282},
numpages = {14},
location = {San Antonio, Texas},
series = {POPL '79}
}

@inproceedings{abstract_interpretation,
author = {Cousot, Patrick and Cousot, Radhia},
title = {Abstract interpretation: a unified lattice model for static analysis of programs by construction or approximation of fixpoints},
year = {1977},
isbn = {9781450373500},
publisher = {Association for Computing Machinery},
address = {New York, NY, USA},
url = {https://doi.org/10.1145/512950.512973},
doi = {10.1145/512950.512973},
booktitle = {Proceedings of the 4th ACM SIGACT-SIGPLAN Symposium on Principles of Programming Languages},
pages = {238–252},
numpages = {15},
location = {Los Angeles, California},
series = {POPL '77}
}

@inproceedings{sea_of_nodes,
author = {Click, Cliff and Paleczny, Michael},
title = {A simple graph-based intermediate representation},
year = {1995},
isbn = {0897917545},
publisher = {Association for Computing Machinery},
address = {New York, NY, USA},
url = {https://doi.org/10.1145/202529.202534},
doi = {10.1145/202529.202534},
booktitle = {Papers from the 1995 ACM SIGPLAN Workshop on Intermediate Representations},
pages = {35–49},
numpages = {15},
location = {San Francisco, California, USA},
series = {IR '95}
}

@misc{colored_egraphs,
      title={Colored E-Graph: Equality Reasoning with Conditions}, 
      author={Eytan Singher and Shachar Itzhaky},
      year={2023},
      eprint={2305.19203},
      archivePrefix={arXiv},
      primaryClass={cs.PL},
      url={https://arxiv.org/abs/2305.19203}, 
}

@INPROCEEDINGS{easter_egg,
  author={Singher, Eytan and Itzhaky, Shachar},
  booktitle={2024 Formal Methods in Computer-Aided Design (FMCAD)}, 
  title={Easter Egg: Equality Reasoning Based on E-Graphs with Multiple Assumptions}, 
  year={2024},
  volume={},
  number={},
  pages={70-83},
  doi={10.34727/2024/isbn.978-3-85448-065-5_13}
}

@article{monotone_frameworks,
author = {Kam, John B. and Ullman, Jeffrey D.},
title = {Monotone data flow analysis frameworks},
year = {1977},
issue_date = {September 1977},
publisher = {Springer-Verlag},
address = {Berlin, Heidelberg},
volume = {7},
number = {3},
issn = {0001-5903},
url = {https://doi.org/10.1007/BF00290339},
doi = {10.1007/BF00290339},
journal = {Acta Inf.},
month = sep,
pages = {305–317},
numpages = {13}
}

@misc{xla,
  title	= {XLA : Compiling Machine Learning for Peak Performance},
  author	= {Amit Sabne},
  year	= {2020}
}

@inproceedings{semantic_sea_of_nodes,
author = {Demange, Delphine and Fern\'{a}ndez de Retana, Yon and Pichardie, David},
title = {Semantic reasoning about the sea of nodes},
year = {2018},
isbn = {9781450356442},
publisher = {Association for Computing Machinery},
address = {New York, NY, USA},
url = {https://doi.org/10.1145/3178372.3179503},
doi = {10.1145/3178372.3179503},
booktitle = {Proceedings of the 27th International Conference on Compiler Construction},
pages = {163–173},
numpages = {11},
location = {Vienna, Austria},
series = {CC '18}
}

@article{labeled_union_find,
author = {Lesbre, Dorian and Lemerre, Matthieu and Ait-El-Hara, Hichem Rami and Bobot, Fran\c{c}ois},
title = {Relational Abstractions Based on Labeled Union-Find},
year = {2025},
issue_date = {June 2025},
publisher = {Association for Computing Machinery},
address = {New York, NY, USA},
volume = {9},
number = {PLDI},
url = {https://doi.org/10.1145/3729298},
doi = {10.1145/3729298},
journal = {Proc. ACM Program. Lang.},
month = jun,
articleno = {195},
numpages = {26}
}

@book{ssa_book,
author = {Rastello, Fabrice},
title = {SSA-based Compiler Design},
year = {2016},
isbn = {1441962018},
publisher = {Springer Publishing Company, Incorporated},
edition = {1st}
}

@inproceedings{tensat,
  title={Equality Saturation for Tensor Graph Superoptimization},
  author={Yichen Yang and Phitchaya Mangpo Phothilimtha and Yisu Remy Wang and Max Willsey and Sudip Roy and Jacques Pienaar},
  eprint={2101.01332},
  booktitle={Proceedings of Machine Learning and Systems},
  year={2021}
}

@misc{co_egraphs,
author = {Zucker, Philip},
title = {Co-Egraphs: Streams, Unification, PEGs, Rational Lambdas},
year = {2024},
month = {July},
day = {29},
url = {https://www.philipzucker.com/coegraph/},
}

@phdthesis{nelson1980,
author = {Nelson, Charles Gregory},
title = {Techniques for program verification},
year = {1980},
publisher = {Stanford University},
address = {Stanford, CA, USA},
note = {AAI8011683}
}

@inproceedings{z3,
author = {De Moura, Leonardo and Bj\o{}rner, Nikolaj},
title = {Z3: an efficient SMT solver},
year = {2008},
isbn = {3540787992},
publisher = {Springer-Verlag},
address = {Berlin, Heidelberg},
booktitle = {Proceedings of the Theory and Practice of Software, 14th International Conference on Tools and Algorithms for the Construction and Analysis of Systems},
pages = {337–340},
numpages = {4},
location = {Budapest, Hungary},
series = {TACAS'08/ETAPS'08}
}

@inproceedings{efficient_e_matching,
author = {Moura, Leonardo and Bj\o{}rner, Nikolaj},
title = {Efficient E-Matching for SMT Solvers},
year = {2007},
isbn = {9783540735946},
publisher = {Springer-Verlag},
address = {Berlin, Heidelberg},
url = {https://doi.org/10.1007/978-3-540-73595-3_13},
doi = {10.1007/978-3-540-73595-3_13},
booktitle = {Proceedings of the 21st International Conference on Automated Deduction: Automated Deduction},
pages = {183–198},
numpages = {16},
location = {Bremen, Germany},
series = {CADE-21}
}

@article{relational_e_matching,
author = {Zhang, Yihong and Wang, Yisu Remy and Willsey, Max and Tatlock, Zachary},
title = {Relational e-matching},
year = {2022},
issue_date = {January 2022},
publisher = {Association for Computing Machinery},
address = {New York, NY, USA},
volume = {6},
number = {POPL},
url = {https://doi.org/10.1145/3498696},
doi = {10.1145/3498696},
journal = {Proc. ACM Program. Lang.},
month = jan,
articleno = {35},
numpages = {22}
}

@inproceedings{eqsat_llvm_tvalid,
author = {Stepp, Michael and Tate, Ross and Lerner, Sorin},
title = {Equality-based translation validator for LLVM},
year = {2011},
isbn = {9783642221095},
publisher = {Springer-Verlag},
address = {Berlin, Heidelberg},
booktitle = {Proceedings of the 23rd International Conference on Computer Aided Verification},
pages = {737–742},
numpages = {6},
location = {Snowbird, UT},
series = {CAV'11}
}

@inproceedings{towards_relational_e_graph,
  title={PLDI: U: Towards a Relational E-graph},
  author={Yihong Zhang},
  year={2022},
  url={https://api.semanticscholar.org/CorpusID:250122313}
}

@inproceedings{mine_dbms,
author = {Min\'{e}, Antoine},
title = {A New Numerical Abstract Domain Based on Difference-Bound Matrices},
year = {2001},
isbn = {3540420681},
publisher = {Springer-Verlag},
address = {Berlin, Heidelberg},
booktitle = {Proceedings of the Second Symposium on Programs as Data Objects},
pages = {155–172},
numpages = {18},
series = {PADO '01}
}

@inproceedings{mine_graph_based,
author = {Min\'{e}, Antoine},
title = {A Few Graph-Based Relational Numerical Abstract Domains},
year = {2002},
isbn = {3540442359},
publisher = {Springer-Verlag},
address = {Berlin, Heidelberg},
booktitle = {Proceedings of the 9th International Symposium on Static Analysis},
pages = {117–132},
numpages = {16},
series = {SAS '02}
}

@article{mine_octagon,
author = {Min\'{e}, Antoine},
title = {The octagon abstract domain},
year = {2006},
issue_date = {March     2006},
publisher = {Kluwer Academic Publishers},
address = {USA},
volume = {19},
number = {1},
issn = {1388-3690},
url = {https://doi.org/10.1007/s10990-006-8609-1},
doi = {10.1007/s10990-006-8609-1},
journal = {Higher Order Symbol. Comput.},
month = mar,
pages = {31–100},
numpages = {70}
}

@inproceedings{pentagons,
author = {Logozzo, Francesco and F\"{a}hndrich, Manuel},
title = {Pentagons: a weakly relational abstract domain for the efficient validation of array accesses},
year = {2008},
isbn = {9781595937537},
publisher = {Association for Computing Machinery},
address = {New York, NY, USA},
url = {https://doi.org/10.1145/1363686.1363736},
doi = {10.1145/1363686.1363736},
booktitle = {Proceedings of the 2008 ACM Symposium on Applied Computing},
pages = {184–188},
numpages = {5},
location = {Fortaleza, Ceara, Brazil},
series = {SAC '08}
}

@article{tvpi,
author = {Simon, Axel and King, Andy},
title = {The two variable per inequality abstract domain},
year = {2010},
issue_date = {March     2010},
publisher = {Kluwer Academic Publishers},
address = {USA},
volume = {23},
number = {1},
issn = {1388-3690},
url = {https://doi.org/10.1007/s10990-010-9062-8},
doi = {10.1007/s10990-010-9062-8},
journal = {Higher Order Symbol. Comput.},
month = mar,
pages = {87–143},
numpages = {57}
}

@inproceedings{singh_octagon,
author = {Singh, Gagandeep and P\"{u}schel, Markus and Vechev, Martin},
title = {Making numerical program analysis fast},
year = {2015},
isbn = {9781450334686},
publisher = {Association for Computing Machinery},
address = {New York, NY, USA},
url = {https://doi.org/10.1145/2737924.2738000},
doi = {10.1145/2737924.2738000},
booktitle = {Proceedings of the 36th ACM SIGPLAN Conference on Programming Language Design and Implementation},
pages = {303–313},
numpages = {11},
location = {Portland, OR, USA},
series = {PLDI '15}
}

@inproceedings{llvm,
author = {Lattner, Chris and Adve, Vikram},
title = {LLVM: A Compilation Framework for Lifelong Program Analysis \& Transformation},
year = {2004},
isbn = {0769521029},
publisher = {IEEE Computer Society},
address = {USA},
booktitle = {Proceedings of the International Symposium on Code Generation and Optimization: Feedback-Directed and Runtime Optimization},
pages = {75},
location = {Palo Alto, California},
series = {CGO '04}
}

@inproceedings{hotspot,
author = {Paleczny, Michael and Vick, Christopher and Click, Cliff},
title = {The java hotspotTM server compiler},
year = {2001},
publisher = {USENIX Association},
address = {USA},
booktitle = {Proceedings of the 2001 Symposium on JavaTM Virtual Machine Research and Technology Symposium - Volume 1},
pages = {1},
numpages = {1},
location = {Monterey, California},
series = {JVM'01}
}

@inproceedings{mlir,
author = {Lattner, Chris and Amini, Mehdi and Bondhugula, Uday and Cohen, Albert and Davis, Andy and Pienaar, Jacques and Riddle, River and Shpeisman, Tatiana and Vasilache, Nicolas and Zinenko, Oleksandr},
title = {MLIR: scaling compiler infrastructure for domain specific computation},
year = {2021},
isbn = {9781728186139},
publisher = {IEEE Press},
url = {https://doi.org/10.1109/CGO51591.2021.9370308},
doi = {10.1109/CGO51591.2021.9370308},
booktitle = {Proceedings of the 2021 IEEE/ACM International Symposium on Code Generation and Optimization},
pages = {2–14},
numpages = {13},
location = {Virtual Event, Republic of Korea},
series = {CGO '21}
}

@inproceedings{smoothe,
author = {Cai, Yaohui and Yang, Kaixin and Deng, Chenhui and Yu, Cunxi and Zhang, Zhiru},
title = {SmoothE: Differentiable E-Graph Extraction},
year = {2025},
isbn = {9798400706981},
publisher = {Association for Computing Machinery},
address = {New York, NY, USA},
url = {https://doi.org/10.1145/3669940.3707262},
doi = {10.1145/3669940.3707262},
booktitle = {Proceedings of the 30th ACM International Conference on Architectural Support for Programming Languages and Operating Systems, Volume 1},
pages = {1020–1034},
numpages = {15},
location = {Rotterdam, Netherlands},
series = {ASPLOS '25}
}

@article{eboost,
  title={e-boost: Boosted E-Graph Extraction with Adaptive Heuristics and Exact Solving},
  author={Yin, Jiaqi and Song, Zhan and Chen, Chen and Cai, Yaohui and Zhang, Zhiru and Yu, Cunxi},
  journal={arXiv preprint arXiv:2508.13020},
  year={2025}
}

@article{treewidth-extract,
author = {Goharshady, Amir Kafshdar and Lam, Chun Kit and Parreaux, Lionel},
title = {Fast and Optimal Extraction for Sparse Equality Graphs},
year = {2024},
issue_date = {October 2024},
publisher = {Association for Computing Machinery},
address = {New York, NY, USA},
volume = {8},
number = {OOPSLA2},
url = {https://doi.org/10.1145/3689801},
doi = {10.1145/3689801},
journal = {Proc. ACM Program. Lang.},
month = oct,
articleno = {361},
numpages = {27}
}

@INPROCEEDINGS{tristate,
  author={Vishwanathan, Harishankar and Shachnai, Matan and Narayana, Srinivas and Nagarakatte, Santosh},
  booktitle={2022 IEEE/ACM International Symposium on Code Generation and Optimization (CGO)}, 
  title={Sound, Precise, and Fast Abstract Interpretation with Tristate Numbers}, 
  year={2022},
  volume={},
  number={},
  pages={254-265},
  doi={10.1109/CGO53902.2022.9741267}
}

@inproceedings{herbie,
author = {Panchekha, Pavel and Sanchez-Stern, Alex and Wilcox, James R. and Tatlock, Zachary},
title = {Automatically improving accuracy for floating point expressions},
year = {2015},
isbn = {9781450334686},
publisher = {Association for Computing Machinery},
address = {New York, NY, USA},
url = {https://doi.org/10.1145/2737924.2737959},
doi = {10.1145/2737924.2737959},
booktitle = {Proceedings of the 36th ACM SIGPLAN Conference on Programming Language Design and Implementation},
pages = {1–11},
numpages = {11},
location = {Portland, OR, USA},
series = {PLDI '15}
}

@inproceedings{cad,
author = {Nandi, Chandrakana and Willsey, Max and Anderson, Adam and Wilcox, James R. and Darulova, Eva and Grossman, Dan and Tatlock, Zachary},
title = {Synthesizing structured CAD models with equality saturation and inverse transformations},
year = {2020},
isbn = {9781450376136},
publisher = {Association for Computing Machinery},
address = {New York, NY, USA},
url = {https://doi.org/10.1145/3385412.3386012},
doi = {10.1145/3385412.3386012},
booktitle = {Proceedings of the 41st ACM SIGPLAN Conference on Programming Language Design and Implementation},
pages = {31–44},
numpages = {14},
location = {London, UK},
series = {PLDI 2020}
}

@article{spores,
author = {Wang, Yisu Remy and Hutchison, Shana and Leang, Jonathan and Howe, Bill and Suciu, Dan},
title = {SPORES: sum-product optimization via relational equality saturation for large scale linear algebra},
year = {2020},
issue_date = {August 2020},
publisher = {VLDB Endowment},
volume = {13},
number = {12},
issn = {2150-8097},
url = {https://doi.org/10.14778/3407790.3407799},
doi = {10.14778/3407790.3407799},
journal = {Proc. VLDB Endow.},
month = jul,
pages = {1919–1932},
numpages = {14}
}

@misc{aegraphs,
author = {Fallin, Chris},
title = {ægraphs: Acyclic E-graphs for Efficient Optimization in a Production Compiler},
year = {2023},
url = {https://pldi23.sigplan.org/details/egraphs-2023-papers/2/-graphs-Acyclic-E-graphs-for-Efficient-Optimization-in-a-Production-Compiler},
}

@InProceedings{semantic_eqsat,
  author =	{Suciu, Dan and Wang, Yisu Remy and Zhang, Yihong},
  title =	{{Semantic Foundations of Equality Saturation}},
  booktitle =	{28th International Conference on Database Theory (ICDT 2025)},
  pages =	{11:1--11:18},
  series =	{Leibniz International Proceedings in Informatics (LIPIcs)},
  ISBN =	{978-3-95977-364-5},
  ISSN =	{1868-8969},
  year =	{2025},
  volume =	{328},
  editor =	{Roy, Sudeepa and Kara, Ahmet},
  publisher =	{Schloss Dagstuhl -- Leibniz-Zentrum f{\"u}r Informatik},
  address =	{Dagstuhl, Germany},
  URL =		{https://drops.dagstuhl.de/entities/document/10.4230/LIPIcs.ICDT.2025.11},
  URN =		{urn:nbn:de:0030-drops-229523},
  doi =		{10.4230/LIPIcs.ICDT.2025.11}
}

@article{rover_hw,
  author   = {Coward, Samuel and Drane, Theo and Constantinides, George A.},
  journal  = {IEEE Transactions on Computer-Aided Design of Integrated Circuits and Systems},
  title    = {ROVER: RTL Optimization via Verified E-Graph Rewriting},
  year     = {2024},
  volume   = {},
  number   = {},
  pages    = {1-1},
  doi      = {10.1109/TCAD.2024.3410154}
}

@article{nelson1979,
author = {Nelson, Greg and Oppen, Derek C.},
title = {Simplification by Cooperating Decision Procedures},
year = {1979},
issue_date = {Oct. 1979},
publisher = {Association for Computing Machinery},
address = {New York, NY, USA},
volume = {1},
number = {2},
issn = {0164-0925},
url = {https://doi.org/10.1145/357073.357079},
doi = {10.1145/357073.357079},
journal = {ACM Trans. Program. Lang. Syst.},
month = oct,
pages = {245–257},
numpages = {13}
}

@inproceedings{polyhedra,
author = {Cousot, Patrick and Halbwachs, Nicolas},
title = {Automatic discovery of linear restraints among variables of a program},
year = {1978},
isbn = {9781450373487},
publisher = {Association for Computing Machinery},
address = {New York, NY, USA},
url = {https://doi.org/10.1145/512760.512770},
doi = {10.1145/512760.512770},
booktitle = {Proceedings of the 5th ACM SIGACT-SIGPLAN Symposium on Principles of Programming Languages},
pages = {84–96},
numpages = {13},
location = {Tucson, Arizona},
series = {POPL '78}
}

@inproceedings{fast_polyhedra,
author = {Singh, Gagandeep and P\"{u}schel, Markus and Vechev, Martin},
title = {Fast polyhedra abstract domain},
year = {2017},
isbn = {9781450346603},
publisher = {Association for Computing Machinery},
address = {New York, NY, USA},
url = {https://doi.org/10.1145/3009837.3009885},
doi = {10.1145/3009837.3009885},
booktitle = {Proceedings of the 44th ACM SIGPLAN Symposium on Principles of Programming Languages},
pages = {46–59},
numpages = {14},
location = {Paris, France},
series = {POPL '17}
}

@article{karr,
author = {Karr, Michael},
title = {Affine relationships among variables of a program},
year = {1976},
issue_date = {June      1976},
publisher = {Springer-Verlag},
address = {Berlin, Heidelberg},
volume = {6},
number = {2},
issn = {0001-5903},
url = {https://doi.org/10.1007/BF00268497},
doi = {10.1007/BF00268497},
journal = {Acta Inf.},
month = jun,
pages = {133–151},
numpages = {19}
}

@article{abstract_extensionality,
author = {Bruni, Roberto and Giacobazzi, Roberto and Gori, Roberta and Garcia-Contreras, Isabel and Pavlovic, Dusko},
title = {Abstract extensionality: on the properties of incomplete abstract interpretations},
year = {2019},
issue_date = {January 2020},
publisher = {Association for Computing Machinery},
address = {New York, NY, USA},
volume = {4},
number = {POPL},
url = {https://doi.org/10.1145/3371096},
doi = {10.1145/3371096},
journal = {Proc. ACM Program. Lang.},
month = dec,
articleno = {28},
numpages = {28}
}

@article{best_ai,
author = {Giacobazzi, Roberto and Ranzato, Francesco},
title = {The Best of Abstract Interpretations},
year = {2025},
issue_date = {January 2025},
publisher = {Association for Computing Machinery},
address = {New York, NY, USA},
volume = {9},
number = {POPL},
url = {https://doi.org/10.1145/3704882},
doi = {10.1145/3704882},
journal = {Proc. ACM Program. Lang.},
month = jan,
articleno = {46},
numpages = {31}
}

@article{hydra,
author = {Mukherjee, Manasij and Regehr, John},
title = {Hydra: Generalizing Peephole Optimizations with Program Synthesis},
year = {2024},
issue_date = {April 2024},
publisher = {Association for Computing Machinery},
address = {New York, NY, USA},
volume = {8},
number = {OOPSLA1},
url = {https://doi.org/10.1145/3649837},
doi = {10.1145/3649837},
journal = {Proc. ACM Program. Lang.},
month = apr,
articleno = {120},
numpages = {29}
}

@software{artifact,
  author       = {Arbore, Russel and
                  Cheung, Alvin and
                  Willsey, Max},
  title        = {Artifact for "Optimism in Equality Saturation"
                   (PLDI 2026)
                  },
  month        = apr,
  year         = 2026,
  publisher    = {Zenodo},
  doi          = {10.5281/zenodo.19581777},
  url          = {https://doi.org/10.5281/zenodo.19581777},
}

\end{document}